\documentclass[traditabstract,longauth]{aa}

\usepackage{natbib}
\usepackage{graphicx}
\usepackage{amssymb}
\usepackage{amsmath}
\usepackage{hyperref}
\usepackage{xspace}
\usepackage{txfonts}

\newcommand{\fermi}{\textsl{Fermi}\xspace}
\newcommand{\lat}{\textsl{Fermi}/LAT\xspace}

\begin{document}
\title{Radio and Gamma-ray Properties of Extragalactic Jets from the TANAMI Sample}

\author{
  \mbox{M.~B\"ock\inst{\ref{affil:mpifr},\ref{affil:remeis},\ref{affil:wuerzburg}}}\and
  \mbox{M.~Kadler\inst{\ref{affil:wuerzburg}}} \and
  \mbox{C.~M\"uller\inst{\ref{affil:remeis},\ref{affil:wuerzburg},\ref{affil:nijmegen}}} \and
  \mbox{G.~Tosti\inst{\ref{affil:perugia}}} \and
  \mbox{R.~Ojha\inst{\ref{affil:nasa_gsfc},\ref{affil:cua},\ref{affil:umbc}}} \and
  \mbox{J.~Wilms\inst{\ref{affil:remeis}}} \and
  \mbox{D.~Bastieri\inst{\ref{affil:padova_inaf},\ref{affil:padova_uni}}} \and
  \mbox{T.~Burnett\inst{\ref{affil:uni_washington}}} \and
  \mbox{B.~Carpenter\inst{\ref{affil:cua}}} \and
  \mbox{E.~Cavazzuti\inst{\ref{affil:roma}}} \and
  \mbox{M.~Dutka\inst{\ref{affil:cua},\ref{affil:nasa_gsfc}}} \and
  \mbox{J.~Blanchard\inst{\ref{affil:tasmania}}} \and
  \mbox{P.~G.~Edwards\inst{\ref{affil:csiro}}} \and
  \mbox{H.~Hase\inst{\ref{affil:bkg}}} \and
  \mbox{S.~Horiuchi\inst{\ref{affil:csiro_canberra}}} \and
  \mbox{D.~L.~Jauncey\inst{\ref{affil:csiro_canberra}}} \and
  \mbox{F.~Krau\ss{}\inst{\ref{affil:remeis},\ref{affil:wuerzburg}}} \and
  \mbox{M.~L.~Lister\inst{\ref{affil:purdue}}} \and
  \mbox{J.~E.~J.~Lovell\inst{\ref{affil:tasmania}}} \and
  \mbox{B.~Lott\inst{\ref{affil:bordeaux}}} \and
  \mbox{D.~W.~Murphy\inst{\ref{affil:jpl}}} \and
  \mbox{C.~Phillips\inst{\ref{affil:csiro}}} \and
  \mbox{C.~Pl\"otz\inst{\ref{affil:bkg}}} \and
  \mbox{T.~Pursimo\inst{\ref{affil:not}}} \and
  \mbox{J.~Quick\inst{\ref{affil:hartrao}}} \and
  \mbox{E.~Ros\inst{\ref{affil:obs_valencia},\ref{affil:dep_valencia},\ref{affil:mpifr}}} \and
  \mbox{G.~Taylor\inst{\ref{affil:unm},\ref{affil:adj_nrao}}} \and
  \mbox{D.~J.~Thompson\inst{\ref{affil:nasa_gsfc}}} \and
  \mbox{S.~J.~Tingay\inst{\ref{affil:swinburne}}} \and
  \mbox{A.~Tzioumis\inst{\ref{affil:csiro}}} \and
  \mbox{J.~A.~Zensus\inst{\ref{affil:mpifr}}}
  }
\institute{
  Max-Planck-Institut f\"ur Radioastronomie, Auf dem H\"ugel 69,
  53121 Bonn, Germany \email{boeck.moritz@gmail.com}\label{affil:mpifr}
  \and
  Dr.\ Karl-Remeis-Sternwarte, Astronomisches Institut der
  Universit\"at Erlangen-N\"urnberg, and Erlangen Centre for
  Astroparticle Physics, Sternwartstra\ss{}e~7, 96049 
  Bamberg, Germany \email{cornelia.mueller@sternwarte.uni-erlangen.de,
    joern.wilms@sternwarte.uni-erlangen.de}\label{affil:remeis}
  \and
  Lehrstuhl f\"ur Astronomie, Universit\"at\ W\"urzburg, Emil-Fischer Str.\ 31,
  97074 W\"urzburg, Germany
  \email{matthias.kadler@astro.uni-wuerzburg.de} \label{affil:wuerzburg}
  \and
  Department of Astrophysics, Institute for Mathematics, Astrophysics and
  Particle Physics, Radboud University Nijmegen, PO Box 9010, 6500 GL,
  Nijmegen, The Netherlands\label{affil:nijmegen}
  \and
  INFN/University of Perugia, 06123 Perugia, Italy
  \email{gino.tosti@pg.infn.it} \label{affil:perugia}
  \and
  NASA Goddard Space Flight Center, Astrophysics Science Division,
  Code~661, Greenbelt, MD 20771, USA \email{roopesh.ojha@nasa.gov} \label{affil:nasa_gsfc}
  \and
  Institute for Astrophysics \& Computational Sciences, Catholic University 
  of America, Washington, DC 20064, USA \label{affil:cua}
  \and
  CRESST/University of Maryland, Baltimore County, 1000 Hilltop
  Circle, Baltimore, MD 21250, USA\label{affil:umbc} 
  \and
  Istituto Nazionale di Fisica Nucleare, Sezione di Padova, I-35131
  Padova, Italy \label{affil:padova_inaf}
  \and
  Dipartimento di Fisica e Astronomia ``G. Galilei'', Universit\`a di
  Padova, I-35131 Padova, Italy \label{affil:padova_uni}
  \and
  University of Washington, Seattle, WA 98195, USA
  \label{affil:uni_washington}
  \and
  Agenzia Spaziale Italiana (ASI) Science Data Center, I-00133 Roma,
  Italy \label{affil:roma}
  \and
  School of Mathematics \& Physics, University of Tasmania, Private
  Bag 37, Hobart, Tasmania 7001, Australia \label{affil:tasmania}
  \and
  CSIRO Astronomy and Space Science, ATNF, PO Box 76 Epping, NSW 1710,
  Australia \label{affil:csiro}
  \and
  Bundesamt f\"ur Kartographie und Geod\"asie, 93444 Bad K\"otzting,
  Germany \label{affil:bkg}
  \and 
  CSIRO Astronomy and Space Science, Canberra Deep Space
  Communications Complex, P.O.\ Box 1035, Tuggeranong, ACT 2901,
  Australia \label{affil:csiro_canberra}
  \and
  Hartebeesthoek Radio Astronomy Observatory, Krugersdorp, 1740, South
  Africa \label{affil:hartrao}
  \and
  Department of Physics, Purdue University, 525 Northwestern Avenue,
  West Lafayette, IN 47907, USA \label{affil:purdue}
  \and
  Centre d'\'Etudes Nucl\'eaires de Bordeaux Gradignan, IN2P3/CNRS,
  Universit\'e Bordeaux 1, BP120, 33175 Gradignan Cedex, France
  \label{affil:bordeaux}
  \and
  Jet Propulsion Laboratory, 4800 Oak Grove Drive, Pasadena, CA 91109
  \label{affil:jpl}
  \and
  Nordic Optical Telescope Apartado 474 E-38700 Santa Cruz de La Palma
  Santa Cruz de Tenerife, Spain \label{affil:not}
  \and
  Observatori Astron\`omic, Universitat de Val\`encia, Parc
  Cient\'{\i}fic, C.\ Catedr\'atico Jos\'e Beltr\'an 2, 46980 Paterna,
  Val\`encia, Spain \label{affil:obs_valencia}
  \and
  Departament d'Astronomia i Astrof\'{\i}sica, Universitat de Val\`encia,
  C.\ Dr.\ Moliner 50, 46100 Burjassot, Val\`encia, Spain
  \label{affil:dep_valencia} 
  \and
  Department of Physics and Astronomy, University of New
  Mexico, Albuquerque NM, 87131, USA \label{affil:unm}
  \and
  Adjunct Astronomer at the National Radio Astronomy Observatory, USA
  \label{affil:adj_nrao}
  \and  
  Centre for Astrophysics and Supercomputing, Swinburne University of
  Technology, P.O.\ Box 218, Hawthorn, VIC 3122, Australia
  \label{affil:swinburne}
  }

\date {Received: --- / Accepted: ---}

\abstract{ Using high-resolution radio imaging with VLBI techniques,
  the TANAMI program has been observing the parsec-scale radio jets of
  southern (declination south of $-30^\circ$) $\gamma$-ray bright AGN
  simultaneously with \textsl{Fermi}/LAT monitoring of their
  $\gamma$-ray emission. We present the radio and $\gamma$-ray
  properties of the TANAMI sources based on one year of
  contemporaneous TANAMI and \lat data. A large fraction (72\%) of the
  TANAMI sample can be associated with bright $\gamma$-ray sources for
  this time range. Association rates differ for different optical
  classes with all BL\,Lacs, 76\% of quasars and just 17\% of galaxies
  detected by the LAT. Upper limits were established on the
  $\gamma$-ray flux from TANAMI sources not detected by LAT. This
  analysis led to the identification of three new \fermi sources whose
  detection was later confirmed. The $\gamma$-ray and radio
  luminosities are related by $L_\gamma \propto
  L_\mathrm{r}^{0.89\pm0.04}$. The brightness temperatures of the
  radio cores increase with the average $\gamma$-ray luminosity, and
  the presence of brightness temperatures above the inverse Compton
  limit implies strong Doppler boosting in those sources. The
  undetected sources have lower $\gamma$/radio luminosity ratios and
  lower contemporaneous brightness temperatures. Unless the
  \lat-undetected blazars are strongly $\gamma$-ray-fainter than the
  \lat-detected ones, their $\gamma$-ray luminosity should not be
  significantly lower than the upper limits calculated here. }


\maketitle

\section{Introduction}
\label{sect:intro}

Blazars are a subset of active galactic nuclei (AGN). They are very
luminous, strongly variable, and show strong polarized emission
\citep{Urry1995}. These properties can be explained by emission from
collimated jets consisting of charged particles moving at relativistic
velocities that are oriented at a small angle to the line of sight and
thus Doppler boosted \citep{Blandford1978,Maraschi1992}. This
explanation is confirmed by the fact that blazars typically exhibit
apparent superluminal motion in the inner radio jet \citep[see,
e.g.,][for an extensive study]{Lister2013}. While detailed
understanding of their emission processes is still a work in progress
\citep[e.g.,][and references
therein]{Ghisellini2009,ghisellini:10a,tavecchio:10a}, a close link
between radio and high-energy emission from blazars is clear
\citep{fossati:98a,Kovalev2009,Ackermann2011}.

Detailed studies of AGN in the MeV-to-GeV energy regime started with
the EGRET detector aboard the \textsl{Compton Gamma Ray Observatory}
\citep[\textsl{CGRO};][]{Thompson1993_egret}, which found that many
blazars are strong $\gamma$-ray emitters \citep[][and references
  therein]{Hartman1992,Mattox1996,Bloom2008}. Radio observations,
particularly using Very Long Baseline Interferometry (VLBI), were
immediately recognized as a particularly useful tool to understand
high-energy emission from blazars. VLBI provides information at the
highest possible resolution and is the only way to measure kinematics
of a blazar jet, including many of the parameters that are essential
inputs to models that seek to explain blazar emission processes
\citep[e.g.,][]{Cohen2007}. Early studies using EGRET and VLBI gave
astronomers the first good glimpse of the high-energy blazar emission
and its connection to radio properties, e.g., compared to
non-detections, EGRET detections had a higher radio flux density and
variability \citep{Impey1996,Tingay2003a}, higher brightness
temperatures \citep{Moellenbrock1996}, more strongly polarized jets
\citep{Lister2005} and larger than average opening angles
\citep{Taylor2007}.

Despite the insights provided into high-energy blazar emission, EGRET
data were limited in many ways, such as by the difficulty of precise
determination of source positions, the very non-uniform sky coverage,
poor temporal sampling, and the limited sensitivity. For that reason
many results were tentative, incomplete, or even inconsistent. Most
obviously, many of the most radio luminous and compact blazars were
not detected. A greatly enhanced successor to EGRET, the Large Area
Telescope (LAT) on board the \textsl{Fermi Gamma-ray Space Telescope}
\citep{Atwood2009}, was launched on 2008 June 11 and commenced regular
observations two months later. \textsl{Fermi}/LAT is a pair conversion
detector of $\gamma$-rays with energies in the range $\sim$20\,MeV to
$>$300\,GeV. The sensitivity of LAT is more than an order of magnitude
higher than that of EGRET. In LAT's sky-survey mode the entire sky is
scanned every 3\,hours, and fairly uniform exposure is obtained within
two months. One of LAT's major scientific goals is to observe the
$\gamma$-ray activity of AGN: to detect, monitor, and characterize
rapidly variable flaring sources. Most of the EGRET detections have
been confirmed by \textsl{Fermi}. Detailed discussions of AGN detected
with \lat are given in ``The First Catalog of Active Galactic Nuclei
Detected by the \textsl{Fermi} Large Area Telescope''
\citep[1LAC;][]{1LAC2010}, and its second 
\citep[2LAC;][]{2LAC2011} and third revisions
\citep[3LAC;][]{3LAC2015}. 

The TANAMI \citep[Tracking Active Galactic Nuclei with Austral
  Milliarcsecond Interferometry;] []{Ojha2010} program is a VLBI
monitoring program targeting AGN jets south of $-30^\circ$
declination. Observations are made at two radio frequencies (8 and
22\,GHz) approximately every two months using the telescopes of the
Australian Long Baseline Array \citep[LBA; e.g.][]{Ojha2004} in
combination with telescopes in Australia (NASA's Tidbinbilla
facility), South Africa, Antarctica, Chile, and New Zealand. The array
has been further expanded by the inclusion of one of the antennas of
the ASKAP (Australian Square Kilometre Array Pathfinder) array and the
Warkworth antenna in New Zealand \citep[see, e.g.,][]{Tzioumis2010}
and the new AuScope antennas at Yarragadee (Western Australia) and
Katherine (Northern Territory) \citep[see, e.g.,][]{Lovell2013}. The
dual frequency nature of the VLBI observations yields spectral index
maps of parsec-scale jet features. The multi-epoch monitoring allows
determination of jet parameters such as jet speed and collimation
angles via tracking of individual jet components. As these critical
parameters cannot be determined by any other observational technique,
the highest possible spatial resolution provided by VLBI is uniquely
important to understanding high-energy emission processes in jets.

Complementary to correlations between the total unresolved radio and
the gamma-ray flux \citep[e.g.,][]{Ackermann2011}, observations by
TANAMI and other groups have shown that the $\gamma$-ray brightness of
AGN as seen by LAT is correlated with VLBI jet properties, such as the
opening angle or jet speed, and found other connections, e.g, between
$\gamma$-ray loudness and synchrotron peak frequency
\citep{Lister2009_gamma_radio,Kovalev2009,Ojha2010,Lister2011,Linford2012b}.
We present the 8.4\,GHz radio and the $\gamma$-ray properties of the
TANAMI sample, as obtained from an analysis of the first 11\,months of
\lat observations, which is the period of time used for the First LAT
Catalog \citep[1FGL;][]{1FGL2010}. The 1FGL data set covers the period
from 2008 August 4 to 2009 July 4. Our analysis is based on the 1FGL
period, because the TANAMI sample has been selected to include
southern AGN detected with \lat in this period of time (details of the
sample selection are provided in Sect.\ \ref{subsect:sample}). We use
only $\gamma$-ray data covering the same period of time as the radio
data. Due to source variability, the usage of non-simultaneous data
would wash out the signal from these correlation studies between the
two energy regimes; \citet{Pushkarev2010} find a radio/$\gamma$-ray
delay on the order of months for the best correlation. At higher radio
frequencies shorter time delays are observed. By studying variability
of individual sources \citet{LeonTavares2011} found that the brightest
$\gamma$-ray emission (seen with \lat) occurs in the rising phase of
millimeter flares.

We present our analysis approach in Sect.~\ref{sect:analysis}. Results
of the analysis are presented in Sect.~\ref{sec:results}, including
the detection statistics and possible non-1FGL $\gamma$-ray
counterparts of TANAMI sources. Section~\ref{sec:individual_sources}
is devoted to a discussion of some individual sources. We discuss our
results in Sect.~\ref{sec:discussion} and end with our conclusions in
Sect.~\ref{sec:conclusions}.

\section{Analysis}\label{sect:analysis}

\subsection{The TANAMI sample}\label{subsect:sample}

The TANAMI sample, which is analyzed here, contains 75 AGN. It is
defined as a combined radio and $\gamma$-ray selected sample that
includes most radio-loud extragalactic jets south of $\delta =
-30^\circ$ that have either been detected at $\gamma$-ray energies or
are considered candidate $\gamma$-ray sources.

The sample of 75 AGN includes the initial TANAMI sample of 43 southern
sources. The latter consists of a radio-selected flux-density-limited
sub-sample (with $S_\mathrm{5GHz}>2$\,Jy), a $\gamma$-ray selected
sub-sample of known and candidate $\gamma$-ray loud jets based on
results of \textsl{CGRO}/EGRET, and sources from special classes, such
as intra-day variable (IDV) and GHz peaked spectrum (GPS) sources. The
sample selection and the radio properties of the first 43 sources are
discussed by \citet{Ojha2010}. During the first months of
\textsl{Fermi} operation, several southern AGN that are candidate
counterparts for LAT sources were added to the initial sample of 43
AGN, resulting in a total of 75 objects in the TANAMI sample. This
sample does not include all 101 AGN in this declination range that
have been detected with \lat in the 1FGL time range.

Based on their optical properties, the TANAMI sample is classified
into quasars (Q), BL\,Lac objects (B), and galaxies (G). These
classifications are based on \citet{Veron2006} and
\citet{Shaw2012,Shaw2013}, with the exception of a few cases where we
have updated the source classification based on newer references (see
Sect.\ \ref{sec:individual_sources}). In total the sample contains 38
quasars, 16 BL\,Lac objects, and 11 radio galaxies. For the remaining
10 sources no classifications were available, because most of them are
faint in the optical.

\subsection{$\gamma$-ray counterparts and upper limits}\label{sec:ul_analysis}

In order to study the radio/$\gamma$-ray connection, we searched for
sources in the TANAMI sample with $\gamma$-ray counterparts in the
1FGL catalog. The associations are based on positional coincidence of
the radio and $\gamma$-ray source positions. The associations used are
consistent with those in the 1FGL catalog and its corresponding AGN
catalog, 1LAC \citep{1LAC2010}.

For TANAMI sources without a counterpart in the 1FGL catalog we
calculated upper limits on the $\gamma$-ray flux at the corresponding
radio positions. The upper limits and the test statistic (TS) of these
sources were obtained by a maximum likelihood analysis
\citep{Cash1979,Mattox1996}, where $\sqrt{\mathrm{TS}}$ is comparable
to the significance in $\sigma$. We used only photons in the
``Source'' class of P7\_V6 events with energies in the range
100\,MeV--100\,GeV for the calculation of upper limits. To minimize
contamination from Earth's limb $\gamma$-rays, photons with zenith
angles greater than $100^\circ$ were removed. The standard \lat
\textit{ScienceTools} software
package\footnote{\href{http://fermi.gsfc.nasa.gov/ssc/data/analysis/documentation/Cicerone/}{http://fermi.gsfc.nasa.gov/ssc/data/analysis/documentation/Cicerone/}}
(version v9r23p1) was used with the ``P7SOURCE\_V6'' set of instrument
response functions\footnote{The 1FGL analysis was done using the
  P6\_V3 events and response functions. We used the 1FGL results
  because the fluxes and spectral indices did not change significantly
  with the P7 analysis, as shown by the 2FGL catalog
  \citep{2FGL2012}.}. The flux, photon index, and TS of each source
were determined by analyzing a Region of Interest (RoI) of $10^\circ$
in radius centered at the radio position. We modeled the LAT point
sources with individual power-law spectra (photon flux $dN/dE =
K(E/E_0)^{-\Gamma}$). The Galactic diffuse background
(\texttt{gal\_2yearp7v6\_v0}) and the isotropic background
(\texttt{iso\_p7v6source}) used in the RoI model, including the
$\gamma$-ray diffuse and residual background of misclassified cosmic
rays, are the recommended versions released and described in more
detail in the documentation available at the \textsl{Fermi} Science
Support
Center\footnote{\href{http://fermi.gsfc.nasa.gov/ssc/data/access/lat/BackgroundModels.html}{http://fermi.gsfc.nasa.gov/ssc/data/access/lat/BackgroundModels.html}}.
The \texttt{xml} source model for the analyzed region has been created
using the 1FGL catalog and a modified version of the tool
\texttt{make2FGLxml.py} contributed by T.\ Johnson, where a source has
been added at the radio position for the upper limit calculation. The
tool creates a source model including all sources in the RoI as well
as sources that are close enough to contribute photons in the RoI due
to the instrument response function of \lat. Model parameters from the
1FGL catalog are used as default values for every source in the model.
We fixed the model parameters of sources outside the RoI to their
catalog values, while parameters of sources inside the RoI were
variable in the modelling process.

Uncertainties in the LAT effective area represent the major source of
systematic error in these results. These uncertainties in the
effective area for the IRFs were evaluated by \citet{Ackermann2012} as
10\% at 100\,MeV, 5\% at 560\,MeV, and 10\% above 10\,GeV, linearly
varying with the logarithm of energy between those values. The
statistical uncertainties exceed these values in all cases. The
reported errors on spectral parameters are 1$\sigma$ uncertainties and
statistical only. Following \citet{Ackermann2012}, we estimate that
the systematic uncertainties are comparable or smaller, $\sim$8\% for
the fluxes and $\sim$0.1 in photon indices.

In a few cases this LAT analysis of the unassociated AGN from the
TANAMI sample using the enhanced sensitivity of the P7 data gave
detections ($\mathrm{TS}\geq 25$) instead of upper limits (see Sect.
\ref{sect:tentative_detections} for more information). For the
non-detected sources upper limits on the $\gamma$-ray flux have been
obtained by fixing the spectral index to $\Gamma=2.4$ and increasing
the flux until a $\Delta$TS of $2.71$ was reached, which yields an
upper limit at the 90\% confidence limit. As this method
underestimates the upper limit for sources with $\mathrm{TS}<1$, the
Bayesian method \citep{Helene1983} has been applied in these cases
\citep[for more information see Sect.~4.4 of][]{1FGL2010}.

Throughout this work, the $\gamma$-ray band 100\,MeV--100\,GeV is
used. For comparisons of $\gamma$-ray properties obtained in analyses
done here with corresponding quantities reported in the LAT catalogs
\citep{2FGL2012}, values such as spectral index and energy flux could
be taken directly from these catalogs because the energy ranges
analyzed were the same. As the integrated flux in the
100\,MeV--100\,GeV band is not given directly in the published
catalogs, we calculated this value using the flux density at the pivot
energy and the spectral index \citep[see, e.g.,][for more
  information]{1FGL2010}. The flux uncertainty in this band has been
obtained using the uncertainties of the spectral index and the flux
density.

\subsection{Radio Analysis}\label{sec:radio_analysis}

The VLBI radio analysis of the TANAMI sources follows
\citet{Ojha2010}, who published results for the initial 43 sources in
the sample. For each source we used the first radio epoch with an
observation date within the \fermi 1FGL period as being representative
of the source's radio flux during the 1FGL period.

We determined brightness temperatures for the radio core at 8.4\,GHz
in the source frame as described by \citet{Ojha2010},
\begin{equation}
T_\mathrm{B} = \frac{2 \ln 2}{\pi k_\mathrm{B}} \;
\frac{S_\mathrm{core} \lambda ^2
  (1+z)}{\theta_\mathrm{maj}\theta_\mathrm{min}}
\end{equation}
where $S_\mathrm{core}$, $\theta_\mathrm{maj}$, and
$\theta_\mathrm{min}$ are the flux density (in Janskys), the
semimajor, and semiminor axis of a two dimensional Gaussian model for
the core in the radio image (in milliarcseconds), $k_\mathrm{B}$ is
the Boltzmann constant, $z$ the redshift of the source, and $\lambda$
the observing frequency. If the size of the fitted model component for
the core emission falls below the resolution limit, we calculated
lower limits for the brightness temperature following
\citet{Kovalev2005}. The sources for which only a lower limit on
$T_\mathrm{B}$ can be given are \object{PKS\,0717$-$432},
\object{PKS\,0812$-$736}, and \object{PKS\,1606$-$667} (see
Sect.\ \ref{sect:tentative_detections}).

For a few sources in the sample, no radio properties are presented
here, because either they were not observed in the 1FGL period or no
radio core could be identified in their VLBI image due to an irregular
morphology. Although these sources do not contribute to the study of
radio and $\gamma$-ray emission, we show their $\gamma$-ray properties
for completeness. 

\subsection{Gamma-ray Luminosities}
\label{sec:gamma_lum}
Assuming the sources have a power-law photon flux spectrum of the form
\begin{equation}
    N_\mathrm{ph}(E) = S_\mathrm{ph} \frac{1-\Gamma}{E_0} \left(
      \left(\frac{E_\mathrm{max}}{E_0}\right)^{1-\Gamma} -
      \left(\frac{E_\mathrm{min}}{E_0}\right)^{1-\Gamma} \right)^{-1}
    \left( \frac{E}{E_0}\right)^{-\Gamma}
\end{equation}
where $S_\mathrm{ph}$ is the measured photon flux in the energy band
from $E_\mathrm{min}$ to $E_\mathrm{max}$ ($E_0$ is only a reference
energy providing a dimensionless base for the non-integer exponent),
and where the photon index $\Gamma\ne 1$. The energy flux in that band
is given by
\begin{equation}
    S_\mathrm{E} = S_\mathrm{ph} \;E_0 \; \frac{1-\Gamma}{2-\Gamma} \;
    \frac{(E_\mathrm{max}/E_0)^{2-\Gamma}-(E_\mathrm{min}/E_0)^{2-\Gamma}}
    {(E_\mathrm{max}/E_0)^{1-\Gamma}-(E_\mathrm{min}/E_0)^{1-\Gamma}}
\end{equation}
for $\Gamma\ne2$.

As the sources in the sample are located at different distances, we
corrected the measured luminosities using a K-correction following
\citet{Ghisellini2009}, i.e.,
\begin{equation}
    L_E = 4\pi d_\mathrm{L}^2 \frac{S_\mathrm{E}}{\left( 1+z\right)^{2-\Gamma}}
\end{equation}
where $z$ is the redshift of the source, $S_\mathrm{E}$ the energy
flux, and $\Gamma$ the photon spectral index in the $\gamma$-ray band.
The luminosity distance, $d_\mathrm{L}$, was calculated assuming a
flat universe with $\Omega_\mathrm{M}=0.27$, $\Omega_\Lambda=0.73$,
and $H_0=71.0\,\mathrm{km}\,\mathrm{s}^{-1}\,\mathrm{Mpc}^{-1}$.

\section{Results}\label{sec:results}

\subsection{Detection Statistics}

The analysis of LAT data revealed different results for different
optical classes of AGN. A total of 54 of the 75 AGN from the TANAMI
sample can be associated with $\gamma$-ray sources from the 1FGL
catalog. Table~\ref{tab:assoc_flux} lists the associated AGN. All of
the BL\,Lac objects (16/16) in the sample were detected in the
$\gamma$-ray regime, and 29 out of 38 quasars were detected, but only
2 out of 11 radio galaxies have strong enough $\gamma$-ray emission to
be detected in the analysed 11 months of LAT data. The low detection
fraction for the radio galaxies is consistent with jet inclination
effects and relativistic beaming as predicted by AGN unification
\citep{Urry1995}. Out of the 10 unclassified sources, 7 are detected
with LAT. We note that the fraction of $\gamma$-ray detections of
certain source classes is biased by our inhomogeneous sample selection
(see Sect.~\ref{subsect:sample}), e.g., sources have been added due to
a LAT detection. The added sources include many unclassified objects,
whereas the radio galaxies were all included in the initial TANAMI
sample.

The two \lat-detected radio galaxies are \object{PKS\,1322$-$428}
(Cen\,A) and \object{PKS\,0521$-$365} (ESO 362-G021). Cen\,A, which is
the closest AGN, has the lowest $\gamma$-ray luminosity of all
associated sources. Gamma-ray emission from its central region and
from the giant radio lobes is observed \citep{Abdo2010_CenAcore,
  Abdo2010_CenAlobe}. Due to Cen\,A's proximity, properties of its jet
can be studied with exceptionally high resolution at sub-parsec scales
using radio VLBI \citep[see,
  e.g.,][]{Tingay2001_CenA,Mueller2011,Mueller2014}. The other
detected radio galaxy, \object{PKS\,0521$-$365}, has been suggested to
be a BL\,Lac object based on the properties of its nucleus \citep[see,
  e.g.,][]{Danziger1979}. It was considered as an example of a
misaligned radio galaxy with an innermost jet on mas-scales oriented
close to the line of sight. Interestingly, \citet{Tingay2002} find
that this source is likely not ``strongly affected by relativistic
boosting''. Further TANAMI radio observations will help to clarify the
nature of this source by obtaining detailed properties of its inner
jet.

\begin{figure}
    \includegraphics[width=\columnwidth]{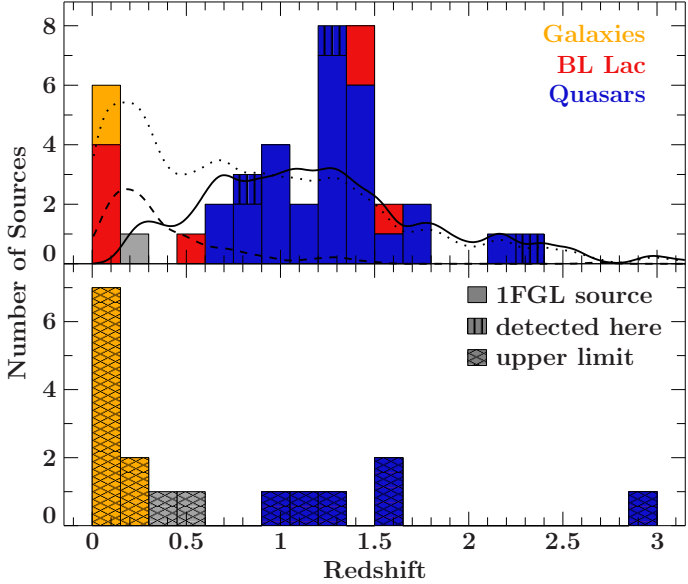}
    \caption{Redshift distribution of LAT detected (top panel) and
      non-detected sources (bottom panel). The lines in the top panel
      represent the redshift distributions of all \lat-detected AGN
      (dotted line), of BL\,Lac objects (dashed line), and quasars
      (solid line) in the 2LAC, which are shown by a kernel density
      estimation (KDE) scaled to the corresponding numbers of TANAMI
      sources. \label{fig:z_lat_det}}
\end{figure}
Figure~\ref{fig:z_lat_det} shows the distribution of redshifts of \lat
$\gamma$-ray detected and undetected sources (those detected in 1FGL
and two detections reported in Sect.~\ref{sect:tentative_detections}.
The third detection does not have a measured redshift). No statistical
differences between the redshift distributions of \lat-detected and
undetected sources of each class are found. A Kolmogorov-Smirnov (KS)
two sample test does not indicate a significant difference between the
redshift distribution of \lat-detected quasars and that of
non-detected quasars. The same applies to the radio galaxies. However,
it has to be noted that for the quasars, as well as for the radio
galaxies, one of the compared distributions includes only a small
number of elements. All radio galaxies in the sample are at low
redshifts, and the majority of them remain undetected. The comparison
of redshift distributions cannot be done with BL\,Lac objects, because
all sources of this type in the TANAMI sample are detected with \lat.
Comparing the redshift distribution of the \lat-detected AGN in the
TANAMI sample with that of all \lat-detected AGN given in 2LAC
indicates slight differences. Relative to 2LAC, the TANAMI sample
contains fewer sources in the moderate redshift regime of around
0.2--0.8. This difference is caused by the selection of the TANAMI
sample. Contrary to the 2LAC ``Clean Sample'' (AGN with Galactic
latitude $|b|>10^\circ$), which includes 395 BL\,Lac objects and 310
quasars, the fraction of quasars is larger in the TANAMI sample with
38 quasars but only 16 BL\,Lac objects. While there is no obvious
difference between the redshift distribution of quasars in TANAMI and
2LAC, the flux-limited sample selection of TANAMI seems to favor the
nearby BL\,Lac objects. The distributions of galaxies in 2LAC are not
shown separately, due to their low number and only small redshifts.

\begin{table*}
    \caption{Properties of TANAMI sources associated with 1FGL
      sources. Luminosities are only available for sources with
      measured redshift. For a few AGN no radio core flux is given,
      because either their morphology did not allow a unique
      identification of a core or they have not been observed within
      the 1FGL time range. \label{tab:assoc_flux}} \scriptsize
    \renewcommand{\arraystretch}{1.2}
    \includegraphics[angle=90,width=\textwidth]{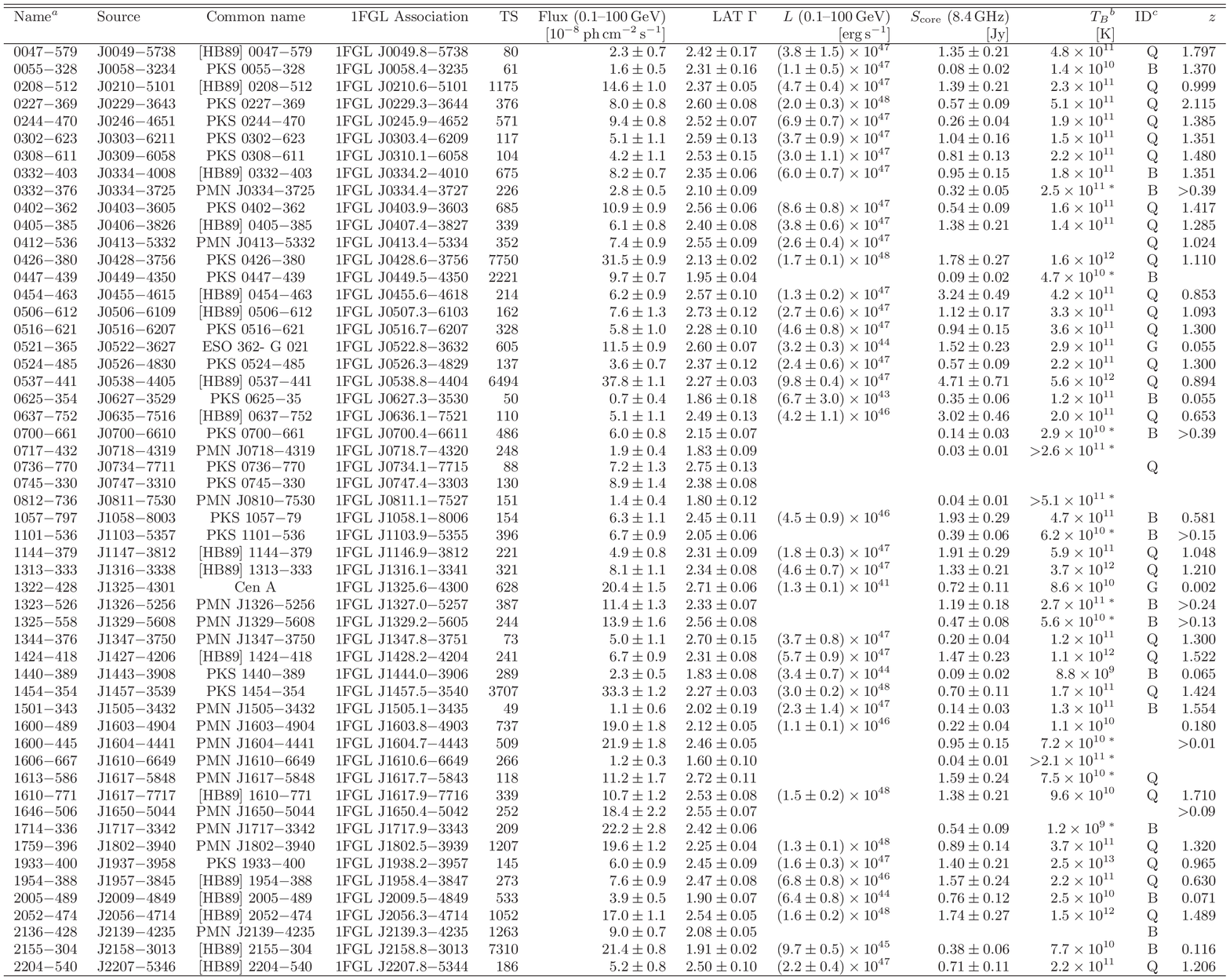}
    $^a$ name in B1950.0 IAU format; $^b$ brightness temperatures
    available in the 1FGL period; for sources without measured
    redshift a lower limit based on $z=0$, or a lower redshift
    limit where available, is indicated with a star ($^\ast$); $^c$
    classifications as described in Sect.\ \ref{subsect:sample}:
    Q: Quasar, B: BL\,Lac, G: galaxy
\end{table*}

\subsection{Possible Non-1FGL Gamma-ray Counterparts}
\label{sect:tentative_detections}
For sample sources without 1FGL $\gamma$-ray counterparts, we
calculated upper limits on the $\gamma$-ray flux as described in
Sect.~\ref{sec:ul_analysis}, using the first 11\,months of \lat data. For
some radio sources, this analysis revealed significant $\gamma$-ray
emission at the radio position. We modeled sources with
$\mathrm{TS} \geq 25$ (Table\ \ref{tab:tent_flux}) and calculated
upper limits for less significant sources (Table\ \ref{tab:ul_flux}).
\begin{table*}
    \caption{Properties of additional $\gamma$-ray detections of
      TANAMI sources in the 1FGL period.\label{tab:tent_flux}} \scriptsize
    \renewcommand{\arraystretch}{1.2}
    \resizebox{\textwidth}{!}{
    \begin{tabular}{llcrrrrrrrrrr}
      \hline
      Name$^a$ & Source & Common name &
      TS & $F$ (0.1--100\,GeV) & LAT $\Gamma$ & $L$
      (0.1--100\,GeV) & $S_\mathrm{core}$ (8.4\,GHz) & $T_\mathrm{B}$$^b$ & ID$^c$
      & $z$ & Sep.$^c$ & Conf95$_\gamma$$^d$
      \\
      & & & &
      [$10^{-8}\,\mathrm{ph}\,\mathrm{cm}^{-2}\,\mathrm{s}^{-1}$] & &
      [$\mathrm{erg}\,\mathrm{s}^{-1}$] 
      & [Jy] & [K] & & & [degree] & [degree] \\
      \hline
      1505$-$496 & J1508$-$4953 &  PMN J1508$-$4953 & 68 & $4.0\pm0.9$ & $2.18\pm0.09$ & $\left(9.8\pm1.8\right)\times10^{46}$ & $0.50\pm0.08$ & $2.2\times10^{12}$ & Q & 0.776 & 0.89 & 0.05\\
      2149$-$306 & J2151$-$3027 &    PKS 2149$-$306 & 61 & $5.1\pm0.9$ & $2.99\pm0.16$ & $\left(2.2\pm0.5\right)\times10^{48}$ & $1.27\pm0.20$ & $1.4\times10^{12}$ & Q & 2.345 & 1.51 & 0.02\\
      2326$-$477 & J2329$-$4730 & [HB89] 2326$-$477 & 27 & $3.0\pm1.0$ &   $3.0\pm0.4$ & $\left(2.2\pm0.9\right)\times10^{47}$ & $0.81\pm0.13$ & $9.7\times10^{10}$ & Q & 1.299 & 0.32 & 0.16\\
      \hline
    \end{tabular}
    }
    \newline
    $^a$ name in B1950.0 IAU format;
    $^b$ brightness temperatures available in the 1FGL period; for
         sources without measured redshift a lower limit based on
	 $z=0$ is indicated with a star ($^\ast$);
    $^c$ classifications as described in Sect.\ \ref{subsect:sample}:
         Q: Quasar, B: BL\,Lac, G: galaxy;
    $^d$ angular separation of the radio position of the TANAMI source
         and the closest 1FGL source;
    $^e$ semimajor axis of the 95\% confidence region of the position
         of the closest 1FGL source
\end{table*}
The detection of additional $\gamma$-ray sources not included in the
1FGL catalog benefited from our usage of the Galactic and
extragalactic background models obtained during the first two years of
\fermi operations, as well as the improved instrument response
functions.

Two quasars (\object{PKS\,2149$-$306} and \object{PKS\,2326$-$477})
and one source with unknown optical counterpart
(\object{PKS\,1505$-$495}) were detected with $\mathrm{TS} \geq 25$,
i.e., they met the detection threshold for the LAT source catalogs.
All three of these detections confirm analysis by the
\textsl{Fermi}/LAT team in the second year \lat-catalog
\citep[2FGL;][]{2FGL2012} and its AGN counterpart, 2LAC
\citep{2LAC2011}.

\begin{table*}
    \caption{Upper limits on the $\gamma$-ray emission of TANAMI
    sources not associated with 1FGL sources. \label{tab:ul_flux}}
    \scriptsize
    \renewcommand{\arraystretch}{1.2}
    \resizebox{\textwidth}{!} {
    \begin{tabular}{llcrrrrrrrrr}
        \hline
        Name$^a$ & Source & Common name &
        TS & $F$ (0.1--100\,GeV) & $L$ (0.1--100\,GeV) &
        $S_\mathrm{core}$ (8.4\,GHz) & $T_\mathrm{B}$$^b$ & ID$^c$ & $z$ &
        Sep.$^d$ & Conf95$_\gamma$$^e$      \\
        & & & &
        [$10^{-8}\,\mathrm{ph}\,\mathrm{cm}^{-2}\,\mathrm{s}^{-1}$] &
        [$\mathrm{erg}\,\mathrm{s}^{-1}$] & 
        [Jy] & [K] & & & [degree] & [degree] \\
        \hline
	0438$-$436 & J0440$-$4333 & [HB89] 0438$-$436 &  2 & $\le1.1$ & $\leq6.8\times10^{47}$ & $0.59\pm0.09$ &          $1.4\times10^{11}$ & Q & 2.863 & 1.69 & 0.02\\
	0518$-$458 & J0519$-$4546 &          PICTOR A & 11 & $\le1.8$ & $\leq2.7\times10^{43}$ & $0.55\pm0.09$ &          $2.7\times10^{10}$ & G & 0.035 & 1.86 & 0.25\\
	0527$-$359 & J0529$-$3555 &  PMN J0529$-$3555 &  0 & $\le1.1$ & $\leq2.1\times10^{45}$ &               &                    &   & 0.323 & 1.51 & 0.10\\
	1104$-$445 & J1107$-$4449 & [HB89] 1104$-$445 &  2 & $\le1.4$ & $\leq1.7\times10^{47}$ & $1.41\pm0.22$ &          $2.7\times10^{11}$ & Q & 1.598 & 0.77 & 0.21\\
	1257$-$326 & J1300$-$3253 &    PKS 1257$-$326 &  2 & $\le2.1$ & $\leq1.4\times10^{47}$ & $0.13\pm0.02$ &          $3.9\times10^{10}$ & Q & 1.256 & 0.35 & 0.13\\
	1258$-$321 & J1301$-$3226 &    ESO 443- G 024 & 18 & $\le2.4$ & $\leq8.3\times10^{42}$ &               &                    & G & 0.017 & 0.56 & 0.13\\
	1333$-$337 & J1336$-$3357 &           IC 4296 &  0 & $\le0.9$ & $\leq1.5\times10^{42}$ & $0.22\pm0.04$ &          $6.2\times10^{9}$ & G & 0.012 & 4.27 & 0.07\\
	1549$-$790 & J1556$-$7914 &     PKS 1549$-$79 &  5 & $\le2.2$ & $\leq7.3\times10^{44}$ & $0.36\pm0.06$ &          $1.6\times10^{10}$ & G & 0.150 & 2.23 & 0.09\\
	1718$-$649 & J1723$-$6500 &          NGC 6328 &  4 & $\le0.8$ & $\leq9.3\times10^{41}$ &               &                    & G & 0.010 & 3.57 & 0.17\\
	1716$-$771 & J1723$-$7713 &    PKS 1716$-$771 &  9 & $\le2.0$ &               & $0.43\pm0.07$ & $9.0\times10^{10}$\,$^\ast$ &   &       & 1.45 & 0.08\\
	1733$-$565 & J1737$-$5634 &     PKS 1733$-$56 &  0 & $\le0.8$ & $\leq1.0\times10^{44}$ & $0.18\pm0.03$ &          $4.9\times10^{10}$ & G & 0.098 & 4.83 & 0.10\\
	1804$-$502 & J1808$-$5011 &  PMN J1808$-$5011 & 23 & $\le2.2$ & $\leq2.9\times10^{47}$ & $0.45\pm0.07$ &          $5.9\times10^{12}$ & Q & 1.606 & 2.28 & 0.15\\
	1814$-$637 & J1819$-$6345 &     PKS 1814$-$63 &  5 & $\le2.1$ & $\leq1.1\times10^{44}$ & $0.32\pm0.05$ &          $6.5\times10^{10}$ & G & 0.063 & 1.95 & 0.18\\
	1934$-$638 & J1939$-$6342 &     PKS 1934$-$63 &  0 & $\le0.9$ & $\leq4.5\times10^{44}$ &               &                    & G & 0.180 & 5.74 & 0.40\\
	2027$-$308 & J2030$-$3039 &    PKS 2027$-$308 &  5 & $\le1.8$ & $\leq1.3\times10^{46}$ & $0.08\pm0.02$ &          $4.7\times10^{9}$ &   & 0.539 & 2.09 & 0.14\\
	2106$-$413 & J2109$-$4110 & [HB89] 2106$-$413 &  0 & $\le0.6$ & $\leq2.4\times10^{46}$ & $1.04\pm0.16$ &          $1.6\times10^{11}$ & Q & 1.058 & 2.94 & 0.12\\
	2152$-$699 & J2157$-$6941 &    ESO 075- G 041 &  5 & $\le1.7$ & $\leq1.6\times10^{43}$ & $0.43\pm0.07$ &          $4.1\times10^{10}$ & G & 0.028 & 2.45 & 0.21\\
	2355$-$534 & J2357$-$5311 & [HB89] 2355$-$534 &  2 & $\le1.7$ & $\leq6.2\times10^{46}$ & $1.53\pm0.23$ &          $5.0\times10^{11}$ & Q & 1.006 & 0.40 & 0.28\\
        \hline
    \end{tabular}
    }
    \newline
    $^a$ name in B1950.0 IAU format;
    $^b$ brightness temperatures available in the 1FGL period; for
         sources without measured redshift a lower limit based on
	 $z=0$ is indicated with a star ($^\ast$);
    $^c$ classifications as described in Sect.\ \ref{subsect:sample}:
         Q: Quasar, B: BL\,Lac, G: galaxy;
    $^d$ angular separation of the radio position of the TANAMI source
         and the closest 1FGL source;
    $^e$ semimajor axis of the 95\% confidence region of the position
         of the closest 1FGL source
\end{table*}

\subsection{Gamma-ray properties of the TANAMI sample}

\begin{figure}
  \includegraphics[width=\columnwidth]{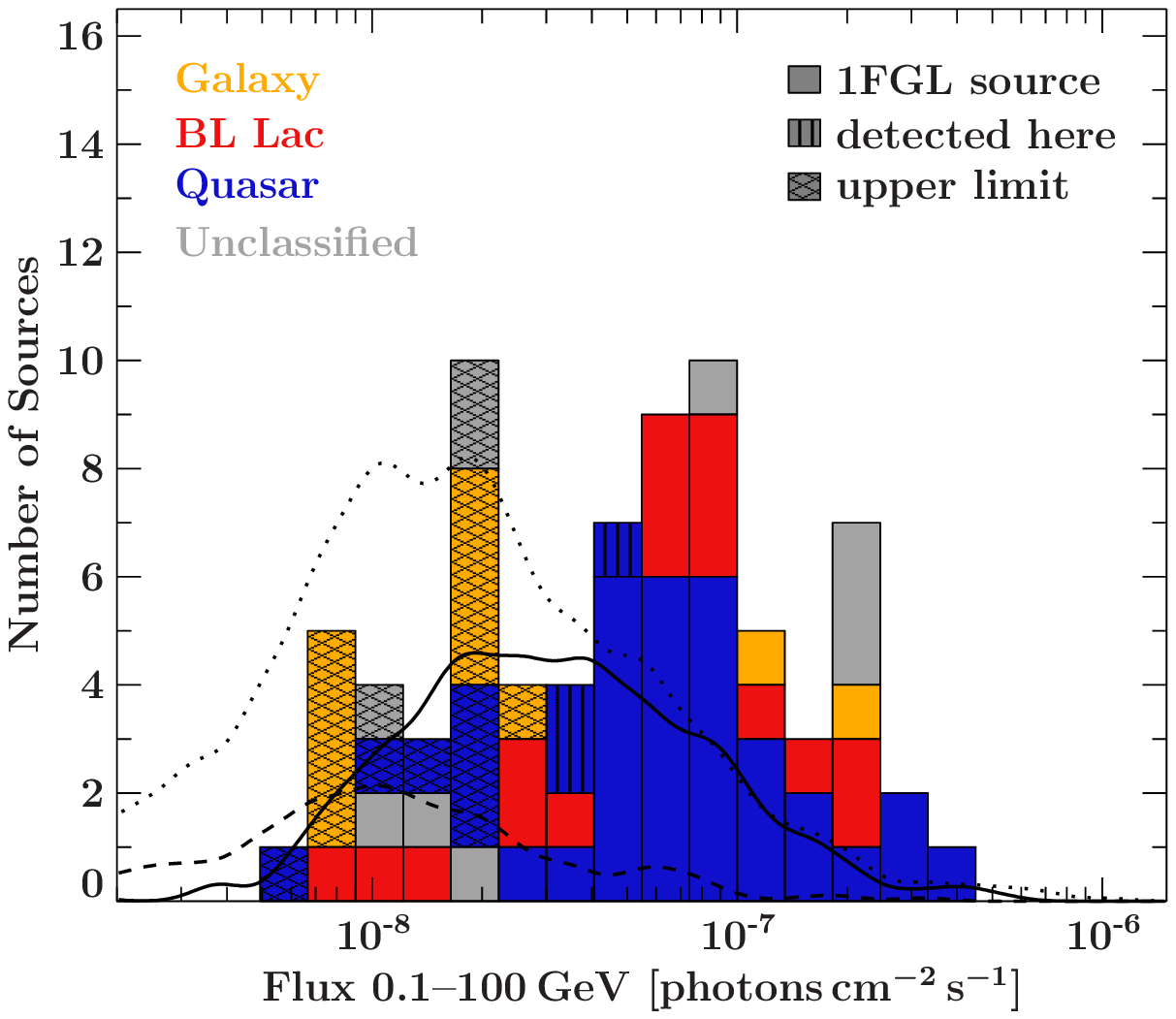}
  \caption{Flux distribution of the \lat-detected sources. For
    comparison the flux distributions of all 2LAC AGN (dotted line),
    the 2LAC quasars (solid line) and the 2LAC BL\,Lacs (dashed line)
    are shown as in Fig.~\ref{fig:z_lat_det}.
    \label{fig:fluxdistr} 
  }
\end{figure}

Figure~\ref{fig:fluxdistr} shows the $\gamma$-ray flux distribution of
the sources in the TANAMI sample. The large number of unclassified
sources mostly results from the addition of new \textsl{Fermi}
detections that are disproportionally fainter and thus less well
studied at other frequencies. The flux and the spectral index are
averaged over the first 11\,months of \fermi science operations. Given
the different source distances it is in general difficult to discern
any clear connection between source type and $\gamma$-ray flux.
The flux distributions of the AGN in 2LAC and in the TANAMI sample are
clearly different, as the sample selection led to a much higher
fraction of bright $\gamma$-ray sources.

\begin{figure}
  \includegraphics[width=\columnwidth]{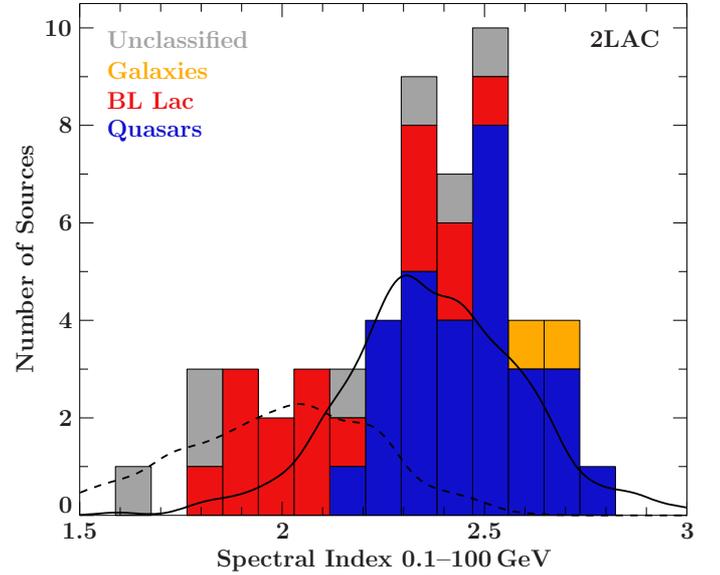}
  \caption{Distribution of the $\gamma$-ray spectral indices $\Gamma$
    of the \lat-detected sources. The lines indicate the distributions
    of spectral indices of quasars (solid line) and BL\,Lac objects
    (dashed line) in the 2LAC (using a KDE as in
    Fig.~\ref{fig:z_lat_det}). \label{fig:indxdistr} }
\end{figure}

The distribution of spectral indices is shown in
Fig.~\ref{fig:indxdistr}. Typical uncertainties are in the range of
$\pm 0.1$ (Table~\ref{tab:assoc_flux}). There is an indication that
BL\,Lacs tend to have, on average, harder spectra than quasars, which
is consistent with earlier \fermi results \citep[][their
  Fig.~12]{1LAC2010}. With a KS two sample test we obtain a
probability of 0.1\% that the $\gamma$-ray spectral indices
of BL\,Lac objects have the same distribution as those of quasars.
This result is consistent with previous studies, in which a relation
between the gamma-ray spectral index and the peak frequency of the
synchrotron component in the spectral energy distribution has been
found. The BL\,Lac objects are generally categorized as low-,
intermediate- and high-synchrotron peaked sources (LSP, ISP, and HSP),
which exhibit softer gamma-ray spectra with decreasing peak frequency
\citep[see, e.g., Fig.~17 of][]{2LAC2011}. This sequence extends to
the quasars with the softest observed gamma-ray spectra.

\begin{figure}
  \includegraphics[width=\columnwidth]{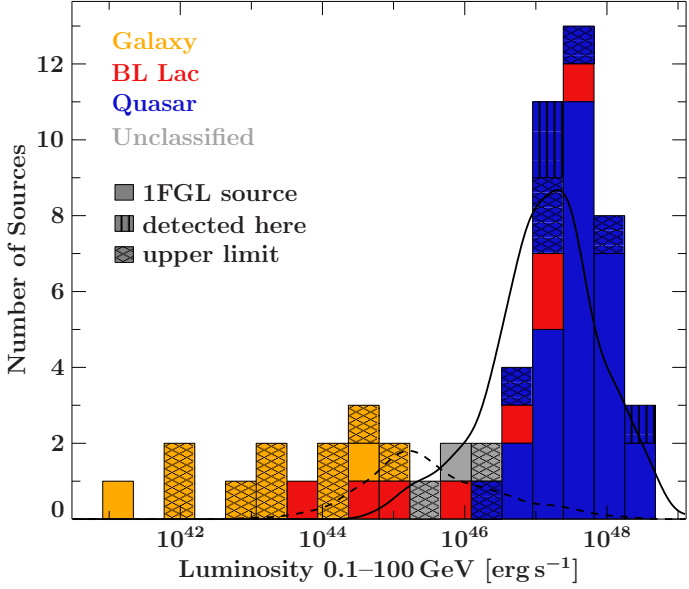}
  \caption{Distribution of $\gamma$-ray luminosities. Radio galaxies
    have low luminosities, BL\,Lac objects are brighter and quasars
    the brightest objects in the sample. For comparison the luminosity
    distributions of quasars and BL\,Lac objects in the 2LAC are shown
    with the solid and dashed line, respectively (using a KDE as in
    Fig.~\ref{fig:z_lat_det}).
    \label{fig:lum_class} }
\end{figure}

The observed $\gamma$-ray luminosities (calculated as described in
Sect.~\ref{sec:gamma_lum} and using a spectral index of $\Gamma=2.4$
for the upper limits) show a clear dependence on the source
classification (Fig.~\ref{fig:lum_class}). Luminosities are low for
radio galaxies, mostly intermediate for BL\,Lac objects, and high for
quasars.
The distribution of upper limits for the luminosity is consistent with
that of the detected sources. The limits for the radio galaxies are
clearly above the luminosity of Cen\,A ($(1.3 \pm 0.1) \times
10^{41}\,\mathrm{erg}\,\mathrm{s}^{-1}$), which is the source with the
lowest measured $\gamma$-ray luminosity in the sample. Upper limits
for the quasars in the sample are not larger than the measured values
but seem to have a similar distribution. A two-sample KS test does not
indicate a significant statistical difference between both
distributions (the probability that upper limits and luminosities of
detected sources have the same underlying distribution is 19\%).

\begin{figure}
  \includegraphics[width=\columnwidth]{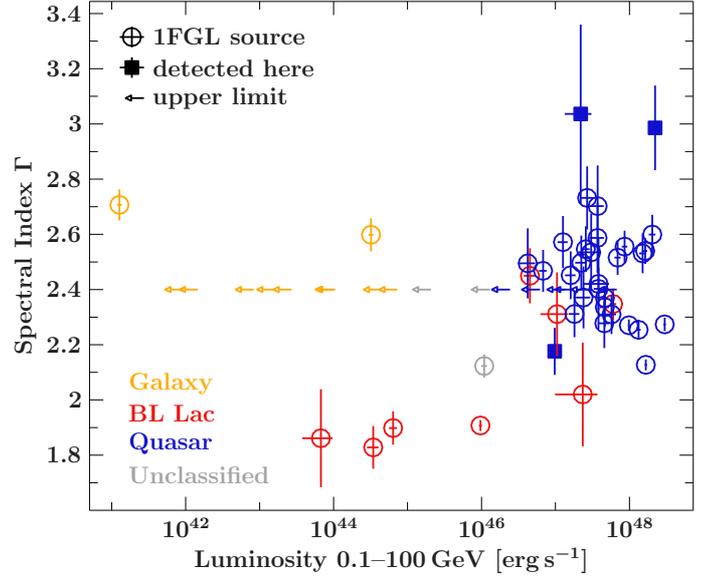}
  \caption{Gamma-ray luminosities and spectral indices for different
    source classes. The upper limits are shown with the average index
    of $\Gamma=2.4$. \label{fig:lum_index}}
\end{figure}

The relation between the $\gamma$-ray luminosity and the spectral
index is shown in Fig.~\ref{fig:lum_index} \citep[this relation for
  all \lat-detected AGN is shown in Fig.~24 of][]{1LAC2010}. The
Pearson correlation coefficient between both quantities (including the
sources detected in the upper limit analysis) is 0.19. This value is
comparable to the value of 0.17 found by \citet{1LAC2010}, who point
out that the correlation might be influenced by instrumental detection
limits and the Malmquist bias.

\subsection{Radio-Gamma-ray-Relations}

\begin{figure}
    \includegraphics[width=\columnwidth]{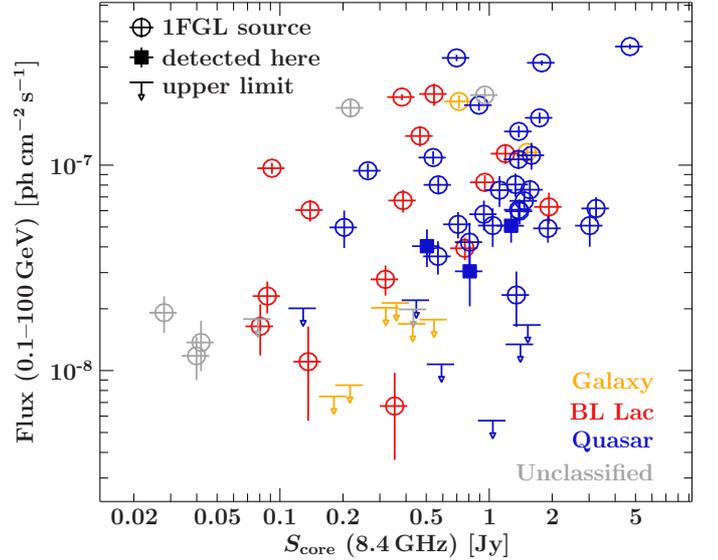}
    \caption{Relation between $\gamma$-ray and radio core flux.}
    \label{fig:gamma_radio_flux}
\end{figure}

\begin{figure}
  \includegraphics[width=\columnwidth]{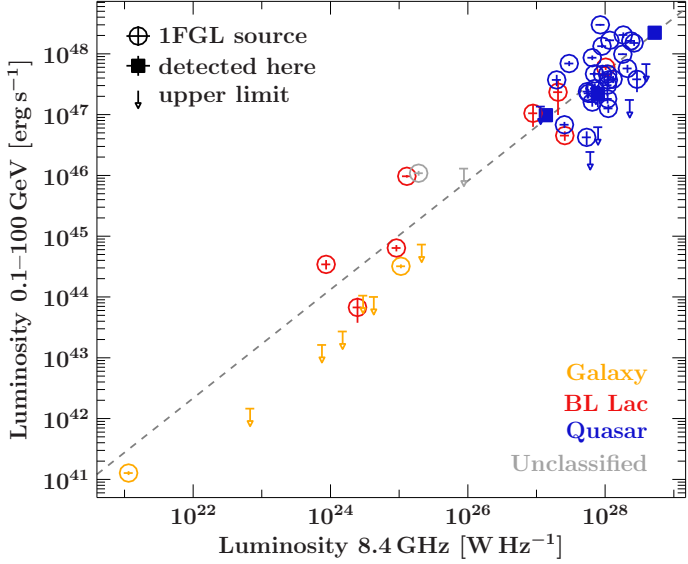}
  \caption{Relation between $\gamma$-ray and radio core luminosity.
    The dashed line shows the fitted power law relation between both
    luminosities. Upper limits tend to be below this relation.}
  \label{fig:gamma_radio_lum}
\end{figure}

Using the results of the radio VLBI imaging, the core flux and the
brightness temperature of the core for each TANAMI source were
calculated (see Sect.\ \ref{sec:radio_analysis}). These radio
properties were compared with the $\gamma$-ray properties.
Figure~\ref{fig:gamma_radio_flux} shows the relation of $\gamma$-ray
flux and the 8.4\,GHz radio core flux density. To quantify
correlations we calculated Kendall's $\tau$ rank correlation
coefficient for censored data \citep[considering determined pairs
only;][]{Helsel2005}. Including the upper limits in this way the
correlation coefficient is 0.29 with a $p$-value of
$3.2\times10^{-5}$, confirming the correlation between these
quantities.
Using the distances and spectral indices of the sources the
luminosities were calculated as described in
Sect.~\ref{sec:gamma_lum}. The relation between the radio and the
$\gamma$-ray luminosity is shown in Fig.~\ref{fig:gamma_radio_lum}.
Fitting a linear function to the logarithm of the luminosities yields
$L_\gamma \propto L_\mathrm{r}^{0.89\pm0.04}$. The clear correlation
between the luminosities is induced by the source distances; thus the
method of \citet{Akritas1996} is used, which yields a partial
correlation coefficient of 0.30 with a $p$-value of
$4.3\times10^{-3}$ between radio and $\gamma$-ray luminosity
given the redshift. These values mainly reflect the correlation
between the fluxes. It is, however, necessary to consider that most of
the sources show significant variability in both energy bands and that
there might be source-dependent time delays between both bands. For
that reason the obtained degree of correlation might be decreased
unless corresponding time periods are used for the observations in
both energy bands.

Figures~\ref{fig:gamma_radio_flux} and \ref{fig:gamma_radio_lum} show
that the $\gamma$-ray/radio brightness ratio is larger for TANAMI
sources that are detected with \fermi than for the \lat-undetected
sources in the sample, where the upper limit is used as the flux.
A significant statistical difference between the distributions of the
luminosity ratios of \lat detected and undetected sources is found. A
two-sample KS test yields a probability of only 1.2\% that both ratios
were drawn from the same underlying distribution.
\begin{figure}
    \includegraphics[width=\columnwidth]{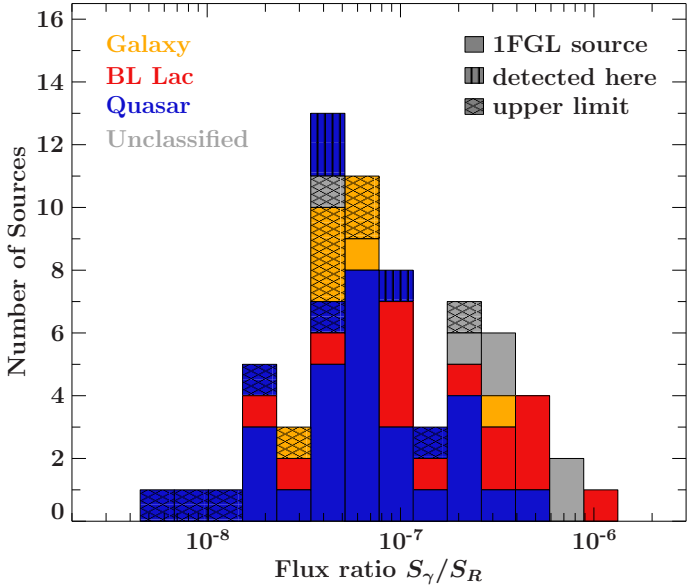}
    \caption{Distribution of the ratio between $\gamma$-ray and radio
      core flux. The stacked histogram is color-coded according to the
      source class. \label{fig:gamma_radio_ratio}}
\end{figure}
The $\gamma$-ray loudness, which is defined here simply as the ratio
of the integral $\gamma$-ray flux to the radio flux density of the
VLBI core, is presented in Fig.~\ref{fig:gamma_radio_ratio} for the
AGN in the sample. The distribution of the $\gamma$-ray loudness
indicates a similar dependence on the source class as the $\gamma$-ray
spectral index distribution: while the quasars are less $\gamma$-ray
loud, the BL\,Lac objects cover a broader range with a slightly higher
averaged $\gamma$-ray loudness. This dependence is consistent with a
shift of the peak frequencies in the SEDs. If the synchrotron peak is
shifted to higher frequencies the flux density in the radio band
decreases, whereas a shift of the high-energy peak in the SED towards
higher frequencies increases the flux in the observed $\gamma$-ray
band. At the same time the spectrum in this band hardens. This
interpretation is strengthened by an observed anti-correlation between
the $\gamma$-ray spectral index, $\Gamma$, and the $\gamma$-ray
loudness as it is defined here (Fig.~\ref{fig:flux_ratio_index}). We
obtain a $p$-value of 1.0\% for a Kendall $\tau$ rank correlation.
\begin{figure}
    \includegraphics[width=\columnwidth]{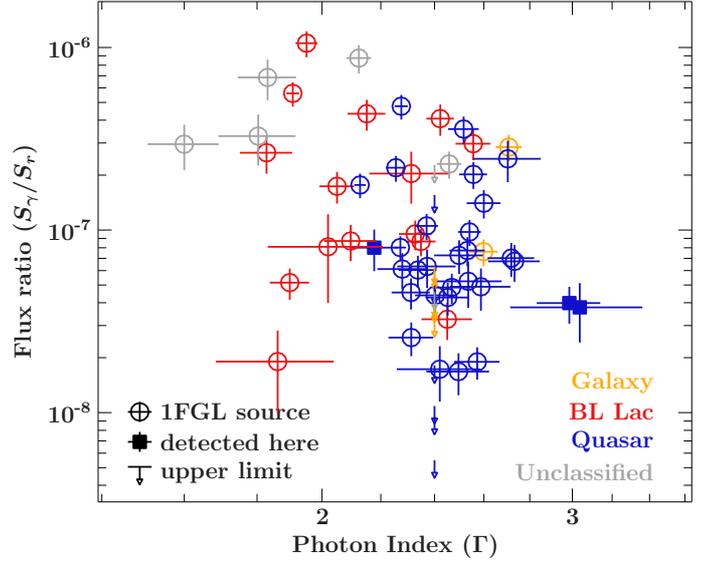}
      \caption{Scatter plot of $\gamma$-ray loudness (defined here as
        a flux ratio) and $\gamma$-ray spectral index.}
      \label{fig:flux_ratio_index}
\end{figure}

\begin{figure}
  \includegraphics[width=\columnwidth]{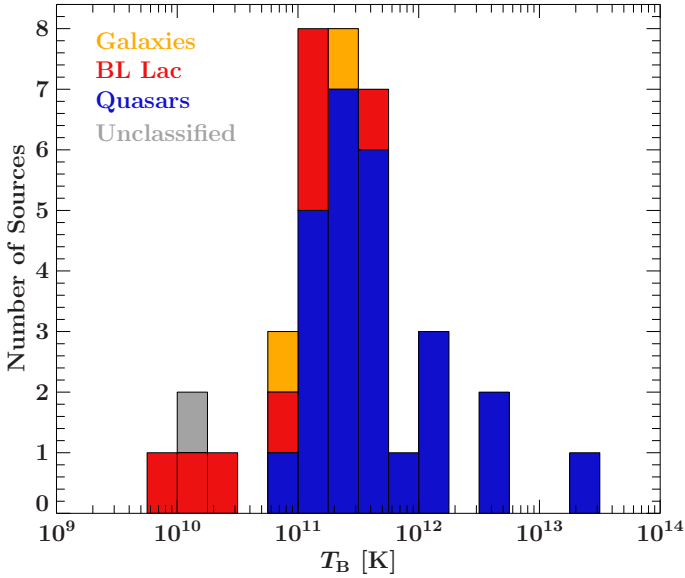}
  \caption{Distribution of brightness temperatures in the TANAMI
    sample.} \label{fig:tb_histogram}
\end{figure}

\begin{figure}
  \includegraphics[width=\columnwidth]{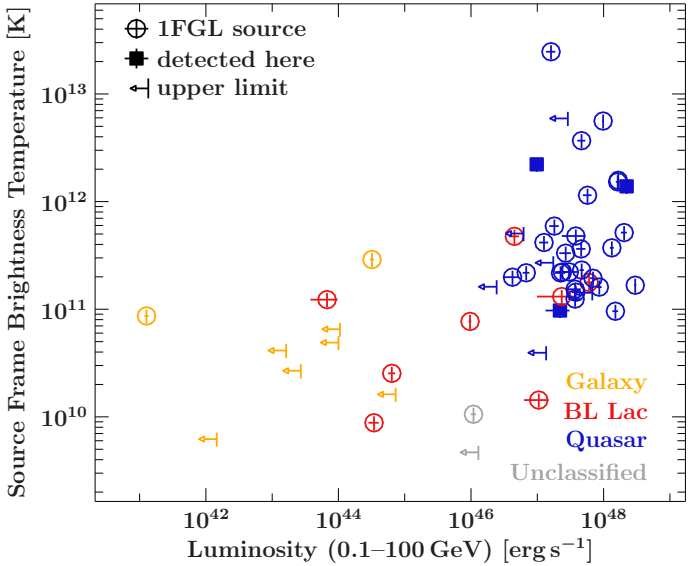}
  \caption{Relation between $\gamma$-ray luminosity and brightness
    temperature of the radio core.}
  \label{fig:gamma_tb}
\end{figure}

Figure~\ref{fig:tb_histogram} shows the distribution of the brightness
temperatures. The distribution is consistent with that obtained from
the first TANAMI observations shown by \citet{Ojha2010}. A broadly
similar distribution of brightness temperatures is also seen in the
MOJAVE survey \citep{Kovalev2009}. The highest observed value is above
$10^{13}$\,K. Due to the theoretical limits on the brightness
temperature in the source frame, such as the inverse Compton limit of
$\sim$$10^{12}$\,K \citep{Kellermann1969}, the larger brightness
temperatures observed here are a clear indication of strong Doppler
boosting \citep[as discussed, e.g., by][]{Tingay2001}.
The relation between brightness temperatures and the $\gamma$-ray
luminosity is shown in Fig.~\ref{fig:gamma_tb}. There is an indication
that the brightness temperature of the radio core increases with
increasing $\gamma$-ray luminosity.
The partial correlation coefficient between $\gamma$-ray luminosity
and brightness temperature, given the redshift, is 0.25 with a
$p$-value of $1.8\times10^{-3}$. \lat undetected sources, of which a
large fraction have a low ratio of $\gamma$-ray to radio luminosity
(Fig.~\ref{fig:gamma_radio_lum}), also tend to have lower brightness
temperatures.

\section{Individual Sources\label{sec:individual_sources}}

In this section we comment on sources for which a classification has
been added and/or the \citet{Veron2006} classification has been
changed. Additionally, sources where our $\gamma$-ray analysis
revealed new results are discussed.

\paragraph{\object{[HB89] 0208$-$512}:}
We use a classification as quasar instead of the BL\,Lac
classification because \citet{Wilkes1986} find a Mg{\sc II} line with
an equivalent width of 18\,\AA.
       
\paragraph{\object{PKS 0302$-$623}:} We use a quasar classification
for this object, as it has properties of a flat spectrum radio quasar
\citep[see, e.g.,][]{Healey2008}.

\paragraph{\object{[HB89] 0332$-$403}:}
The redshift of this source, which is by far the most luminous BL\,Lac
in our sample, is difficult to determine. We use $z=1.351$, which is
based on a single weak Mg~\textsc{ii} emission line
\citep{Bergeron2011} and is consistent with the photometric redshift
of this source \citep[][$z=1.47^{+0.11}_{-0.12}$]{Rau2012}. If the
source classification as a BL\,Lac is correct, \object{PKS 0332$-$403}
would be one of only two BL\,Lac objects with $z>1.3$ in the \fermi
2FGL sample \citep{Rau2012}.

We note, however, that the classification of this source as a BL\,Lac
by \citet{Veron2006} seems not to be very secure. It seems to go back
to \citet{Impey1990}, who found that \object{PKS 0332$-$403} has a
high degree of polarization (14.7\%), above their 2.5\% for
classifying sources as blazars, but little further work appears to
have been done on classifying \object{PKS 0332$-$403}. The more
specialized catalog of BL\,Lacs by \citet{Padovani1995} did not
contain \object{PKS 0332$-$403}. The updated version of that catalog,
Version 4.2 of the Roma-BZCAT catalog \citep{Massaro2009}, lists
\object{PKS 0332$-$403} as only a ``BL\,Lac candidate''.
\citet{Tornianinen2008} and \citet{Tornikoski2001} are more
conservative and list \object{PKS 0332$-$403} as a highly polarized
quasar and possibly a GPS source.

\paragraph{\object{[HB89] 0438$-$436}:}
This is a very bright and luminous radio quasar at high redshift
($z=2.863$) that was not detected in the first 11\,months of \lat
data. The upper limit on the luminosity, given by the flux upper limit
and the distance, is, however, in the range of a relatively luminous
quasar.

\paragraph{\object{PKS 0447$-$439}:}
We do not use a redshift for this source, because it is uncertain and
some of its estimates are even contradictory, as discussed by
\citet{Pita2014}.

\paragraph{\object{ESO 362-G021} (0521$-$365):}
Instead of a BL\,Lac classification we use a galaxy classification
based on the presence of strong emission lines \citep[see,
  e.g.,][]{Falomo1994}.

\paragraph{\object{PKS 0625$-$354}} This source has a large-scale FR-I radio
morphology, but its optical spectrum indicates a BL\,Lac
classification \citep{Wills2004}, which we use in this work.
\object{PKS\,0625$-$354} is one of five misaligned radio galaxies
discussed in the 1LAC 
\citep{1LAC2010}. The TANAMI mas-scale image shows a single-sided jet
with a strong core component and is thus consistent with the inner
parsec-scale jet being oriented close to the line of sight.

\paragraph{\object{PKS 0745$-$330}} has a nearby 1FGL source, but the
radio position is slightly outside the positional 95\% confidence
region of the 1FGL source indicating that it might not be the correct
counterpart. In the 2FGL catalog \citep[2FGL;][]{2FGL2012} the
separation is, however, smaller and the radio and the 2FGL sources are
associated. As the 1FGL and 2FGL sources are clearly associated, we
use the 1FGL as counterpart for \object{0745$-$330} here. Testing the
other case as well, in which the $\gamma$-ray source is not the
correct counterpart for \object{0745$-$330}, the modeling of a
$\gamma$-ray source at the radio position in addition to the nearby
1FGL source yields a TS of 12.7 and a flux of
$\le0.6\times10^{-7}\,\mathrm{ph}\,\mathrm{cm}^{-2}\,\mathrm{s}^{-1}$.

\paragraph{\object{PKS 1057$-$79}:} \citet{Sbarufatti2009}, who
measured the redshift of $z = 0.581$ of this source, list \object{PKS
  1057$-$79} as a BL\,Lac object but suggest classifying it as a
broad-line AGN based on the observed emission lines. In our work we
use the BL\,Lac classification. The $\gamma$-ray luminosity and the
spectral index of this object are in between the typical values for
quasars and BL\,Lacs.

\paragraph{\object{ESO 443-G024} (1258$-$321):}
This is the brightest galaxy in the cluster \object{ACO\,3537}
\citep{Hudson2001}, but has blazar properties as well. With a TS below
25 the source is not detected by \lat. Modeling the source yields a TS
of 18.2, a flux of
$(1.6\pm0.8)\times10^{-8}\,\mathrm{ph}\,\mathrm{cm}^{-2}\,\mathrm{s}^{-1}$,
and a spectral index $2.33\pm0.22$.

\paragraph{\object{PKS 1440$-$389}:} \object{PKS\,1440$-$389} has a very hard
$\gamma$-ray spectrum with spectral index of $1.83\pm0.08$. A BL\,Lac
classification is likely \citep[see, e.g.,][]{Jackson2002,Mao2011}.
The TANAMI VLBI image reveals a weak radio core with
$S_\mathrm{Core}\sim 50$\,mJy and an extension to the southwest.

\paragraph{\object{PMN J1508$-$4953}:}
This source, which was not in 1FGL, was detected in our analysis.
Modeling a $\gamma$-ray source at the radio position of that source
yields a TS of 68, a flux of
$(4.0\pm0.9)\times10^{-8}\,\mathrm{ph}\,\mathrm{cm}^{-2}\,\mathrm{s}^{-1}$,
and a spectral index $2.18\pm0.09$. These values are consistent with
the source's counterpart in the 2FGL catalog
(\object{2FGL\,J1508.5$-$4957}). The properties obtained during the
first two years of \fermi operation are $\mathrm{TS}=70$, a flux of
$S_\mathrm{ph} = (5.5 \pm0.9)\,10^{-8}\,\mathrm{ph}\,\mathrm{cm}^{-2}\,\mathrm{s}^{-1}$ and a
spectral index of $2.61\pm0.09$.

\paragraph{\object{PKS 2149$-$306}:} This source is clearly detected in
the $\gamma$-ray data with a TS of 61, a flux of
$(5.1\pm0.9)\times10^{-8}\,\mathrm{ph}\,\mathrm{cm}^{-2}\,\mathrm{s}^{-1}$,
and a very soft spectral index of $2.99\pm0.16$. With this value, the
source has one of the steepest gamma-ray spectra of all AGN detected
with \lat. The source is included in the 2FGL catalog as
\object{2FGL\,J2151.5$-$3021} and has the following properties
averaged over the first two years of \fermi: $\mathrm{TS}=168$,
$S_\mathrm{ph} = (6.4 \pm
0.7)\times10^{-8}$$\,\mathrm{ph}\,\mathrm{cm}^{-2}\,\mathrm{s}^{-1}$,
and a spectral index of $3.00\pm0.09$.

\paragraph{\object{[HB89] 2326$-$477}:}
is detected with a TS of 27, a flux of
$(3.0\pm1.0)\times10^{-8}\,\mathrm{ph}\,\mathrm{cm}^{-2}\,\mathrm{s}^{-1}$,
and a spectral index of $3.0\pm0.4$. The source is included in 2FGL as
\object{2FGL\,J2329.7$-$4744} with $\mathrm{TS}=25$ and a flux of
$S_\mathrm{ph} =
(1.6\pm0.7)\times10^{-8}\,\mathrm{ph}\,\mathrm{cm}^{-2}\,\mathrm{s}^{-1}$.
The spectral index is $2.58\pm0.22$. With a flux of
$(7.7\pm2.0)\times10^{-8}\,\mathrm{ph}\,\mathrm{cm}^{-2}\,\mathrm{s}^{-1}$
this source was brightest in the tenth of the 24 monthly bins in the
2FGL catalog. The variability explains the slightly higher TS and flux
average in the 1FGL period compared to the full 2FGL period.

\section{Discussion}\label{sec:discussion}

We present $\gamma$-ray properties of the TANAMI sample based on the
data obtained with \lat during its first 11\,months of operation. A
total of 54 out of the 75 AGN from this sample can be associated with
$\gamma$-ray sources from the 1FGL catalog. All BL\,Lac objects
(16/16) and a large fraction of quasars
(29/38) are detected in the $\gamma$-ray regime,
whereas from the 11 radio galaxies only the closest one (Cen\,A) and
one with indications of being a BL\,Lac object
(\object{PKS\,0521$-$365}) were detected with \lat. The low number of
$\gamma$-ray detected radio galaxies is consistent with 1LAC and 2LAC
where the dominating fraction of \lat-detected AGN are blazars. In
2LAC the ``clean'' sample of 886 AGN includes 395 BL\,Lac objects and
310 quasars but only 8 misaligned AGN \citep{2LAC2011}. There does not
appear to be a significant difference in the distribution of redshifts
for \lat detected and non-detected sources as was also found for the
MOJAVE sample \citep{Lister2011}.

For the 21 AGN without $\gamma$-ray counterparts we presented upper
limits on the $\gamma$-ray flux. In three cases
(\object{PKS\,1505$-$496}, \object{PKS\,2149$-$306},
\object{PKS\,2326$-$477}) the TS was high enough to indicate a
detection. All of these three sources are included as detections in
the 2FGL catalog. We note that the upper limits on $\gamma$-ray
luminosity for quasars in the sample are comparable to the measured
values, and both the limits and the values have a similar
distribution. This suggests that the luminosity of the undetected
quasars will be comparable to their upper limits. Alternatively, there
could be a class of fainter (at $\gamma$-ray energies) quasars in the
sample.

A weak correlation between radio and $\gamma$-ray fluxes, as well as
an analogous partial correlation between the luminosities, of the
sources in the TANAMI sample has been found in the first 11\,months of
\fermi science operations (see Figs.~\ref{fig:gamma_radio_flux}
and~\ref{fig:gamma_radio_lum}). Using EGRET data, \citet{Bloom2008}
found a correlation between the $\gamma$-ray and the radio luminosity
at 8.4\,GHz of $L_\gamma \propto L_r^{ 0.77 \pm 0.03}$. They could
reproduce this relation using a synchrotron self-Compton model, but
not with an external Compton model. Studies with \lat data have shown
that the $\gamma$-ray flux correlates well with compact (parsec scale)
radio flux \citep{Kovalev2009}. For the TANAMI sample we find a
relation of $L_\gamma \propto L_r ^{0.89\pm0.04}$. A detailed
interpretation of such a relation is difficult, because the emission
in both bands is variable on different time scales and thus the ratio
between the radio and $\gamma$-ray luminosity is not even constant for
individual sources. With a longer set of contemporaneous observations,
this problem could be addressed by searching for delays, or specifying
well-justified averaging times. A perfect correlation between radio
and $\gamma$-ray luminosity is, however, not expected, as it can be
easily weakened, e.g., by different Doppler boosting in the radio and
$\gamma$-ray regime. Different boosting between both energy regimes
can for example originate from different Lorentz factors in the case
of separated emission regions, but also from different spectral slopes
in the radio and $\gamma$-ray regimes. For that reason, different
angles to the line of sight of different sources in a sample decrease
the observed correlation between the fluxes
\citep[e.g.,][]{Lister2007}. Varying contributions of external
Comptonization will scatter the radio $\gamma$-ray relation further.
The fact that only weak correlations are found between the average
radio and gamma-ray flux agrees with correlated variability being 
observed only rarely in both bands \citep[see, e.g.,][]{Max-Moerbeck2013}.

We find that BL\,Lacs in our sample tend to have harder spectra than
quasars. Early \fermi results as well as the 2LAC show $\gamma$-ray
loud quasars have soft spectra while the BL\,Lacs have a diverse range
of spectral indices, where the gamma-ray spectra soften from HSP over
ISP to LSP BL\,Lac objects \citep{2LAC2011,Lott2012}. The ratio of
$\gamma$-ray and radio flux, which is used here to characterize the
$\gamma$-ray loudness, is anti-correlated with the spectral index in
the $\gamma$-ray band, which is consistent with the above-mentioned
dependence on the synchrotron peak frequency.

Most $\gamma$-ray upper limits for undetected sources tend to be
smaller than required for fitting to the average radio-$\gamma$-ray
relation in the sample, i.e., many of the \lat-undetected AGN are less
$\gamma$-ray-loud than the detected sources. However, it has to be
noted that the derived $\gamma$-ray loudness can be influenced by
variability. As shown by \citet[][their Figs.\ 11 and 19]{1LAC2010},
most AGN show strong variability in the $\gamma$-ray regime. Flares
can strongly increase the measured average flux. Additionally the
sources are variable in the radio regime. Considering possible
emission delays between the bands, the selection of corresponding time
intervals is necessary for a better correlation. \citep[find delays of
about 1.2\,months in the source frame]{Pushkarev2010}. Further
temporal studies, including a search for delays and a comparison of
the jet speeds with $\gamma$-ray properties will be carried out with
more TANAMI epochs over a longer time period.

We find indications that $\gamma$-ray luminous AGN in the TANAMI
sample have larger radio core brightness temperatures than
$\gamma$-ray fainter sources. A comparison of brightness temperatures
of strong EGRET sources and EGRET-undetected sources did not show this
relation \citep[][who used mainly observations at lower
frequencies]{Tingay1998}. A relation between brightness temperature
and $\gamma$-ray brightness has been found in the MOJAVE sample 
\citep{Kovalev2009, Lister2011}.

\section{Conclusions}\label{sec:conclusions}

The radio and $\gamma$-ray properties of the TANAMI AGN sample were
investigated using data obtained during the first 11 months \fermi
operations. Over $70\%$ of the sample had already been detected by
\lat and the rates of detection for quasars and radio galaxies are
consistent with that found in other samples and with studies using
more $\gamma$-ray data.

For those TANAMI sources not in the 1FGL list, an upper limit analysis
was performed. Three new $\gamma$-ray sources were significantly
detected by this analysis. The luminosities of the rest of the
undetected quasars are likely to be close to the upper limits reported
here. The undetected sources have lower $\gamma$-ray-to-radio
luminosity ratio and lower brightness temperatures which fits the
picture of Doppler boosting playing a dominant role in determining the
$\gamma$-ray state of an AGN.

A relation between $\gamma$-ray and radio flux was apparent and the
brightness temperatures of radio cores were found to scale with the
$\gamma$-ray luminosity. Some sources have brightness temperatures
well above the inverse Compton limit suggesting strong Doppler
boosting.

Similar studies of the TANAMI sample will be made using LAT data for
different time ranges to tease out variations on different timescales.
As enough epochs of data are now becoming available for most of the
TANAMI sample, future studies will include VLBI kinematics. The TANAMI
team has upcoming observations with Gemini South, which should allow
optical identifications and redshift measurements for the large
fraction of TANAMI sources (mostly fainter and poorly studied new \lat
detections) that do not have them, significantly improving our
statistics.

\acknowledgements{ This research has been partially funded by the
  Bundesministerium f\"ur Wirtschaft und Technologie under Deutsches
  Zentrum f\"ur Luft- und Raumfahrt grant number 50OR0808. The Long
  Baseline Array is part of the Australia Telescope which is funded by
  the Commonwealth of Australia for operation as a National Facility
  managed by CSIRO. E.R.\ acknowledges partial support by the Spanish
  MINECO grants AYA2009-13036-C02-C02 and AYA2012-38491-C02-01, by the
  Generalitat Valenciana grant Prometeo 2009/104, and by the COST
  action MP0905 ``Black Holes in a Violent Universe''. This research
  was funded in part by NASA through Fermi Guest Investigator grant
  NNH09ZDA001N (proposal number 31263) and grant NNH10ZDA001N
  (proposal number 41213). This research was supported by an
  appointment to the NASA Postdoctoral Program at the Goddard Space
  Flight Center, administered by Oak Ridge Associated Universities
  through a contract with NASA. 
  
  The \lat Collaboration acknowledges generous ongoing support from a
  number of agencies and institutes that have supported both the
  development and the operation of the LAT as well as scientific data
  analysis. These include the National Aeronautics and Space
  Administration and the Department of Energy in the United States,
  the Commissariat \`a l'\'Energie Atomique and the Centre National de
  la Recherche Scientifique / Institut National de Physique
  Nucl\'eaire et de Physique des Particules in France, the Agenzia
  Spaziale Italiana and the Istituto Nazionale di Fisica Nucleare in
  Italy, the Ministry of Education, Culture, Sports, Science and
  Technology (MEXT), High Energy Accelerator Research Organization
  (KEK) and Japan Aerospace Exploration Agency (JAXA) in Japan, and
  the K.A.~Wallenberg Foundation, the Swedish Research Council and
  the Swedish National Space Board in Sweden. Additional support for
  science analysis during the operations phase is gratefully
  acknowledged from the Istituto Nazionale di Astrofisica in Italy and
  the Centre National d'\'Etudes Spatiales in France.
  
  This research has made use of NASA's Astrophysics Data System
  Bibliographic Services. This research has made use of the NASA/IPAC
  Extragalactic Database (NED) which is operated by the Jet Propulsion
  Laboratory, California Institute of Technology, under contract with
  the National Aeronautics and Space Administration. We thank John
  E.\ Davis for the development of the \textsc{SLxfig} module used to
  prepare the figures in this paper. This research has made use of
  ISIS functions provided by ECAP/Remeis observatory and MIT
  (\href{http://www.sternwarte.uni-erlangen.de/isis/}{http://www.sternwarte.uni-erlangen.de/isis/}). We thank Davide
  Donato and Seth Digel for their very helpful comments. }


\begin{thebibliography}{}

\bibitem[\protect\astroncite{Abdo et~al.}{2010a}]{Abdo2010_CenAcore}
Abdo A.A., Ackermann M., Ajello M., et~al., 2010a, ApJ 719, 1433

\bibitem[\protect\astroncite{Abdo et~al.}{2010b}]{Abdo2010_CenAlobe}
Abdo A.A., Ackermann M., Ajello M., et~al., 2010b, Science 328, 725

\bibitem[\protect\astroncite{Abdo et~al.}{2010c}]{1FGL2010}
Abdo A.A., Ackermann M., Ajello M., et~al., 2010c, ApJS 188, 405 {(1FGL)}

\bibitem[\protect\astroncite{Abdo et~al.}{2010d}]{1LAC2010}
Abdo A.A., Ackermann M., Ajello M., et~al., 2010d, ApJ 715, 429 {(1LAC)}

\bibitem[\protect\astroncite{Ackermann et~al.}{2012}]{Ackermann2012}
Ackermann M., Ajello M., Albert A., et~al., 2012, ApJS 203, 4

\bibitem[\protect\astroncite{Ackermann et~al.}{2011a}]{Ackermann2011}
Ackermann M., Ajello M., Allafort A., et~al., 2011a, ApJ 741, 30

\bibitem[\protect\astroncite{Ackermann et~al.}{2011b}]{2LAC2011}
Ackermann M., Ajello M., Allafort A., et~al., 2011b, ApJ 743, 171 {(2LAC)}

\bibitem[\protect\astroncite{Ackermann et~al.}{2015}]{3LAC2015}
Ackermann M., Ajello M., Atwood W.B., et~al., 2015, ApJ  in press {(3LAC;
  arXiv:1501.06954)}

\bibitem[\protect\astroncite{Akritas \& Siebert}{1996}]{Akritas1996}
Akritas M.G., Siebert J.,  1996, MNRAS 278, 919

\bibitem[\protect\astroncite{Atwood et~al.}{2009}]{Atwood2009}
Atwood W.B., Abdo A.A., Ackermann M., et~al., 2009, ApJ 697, 1071

\bibitem[\protect\astroncite{Bergeron et~al.}{2011}]{Bergeron2011}
Bergeron J., {Boiss{\'e}} P., {M{\'e}nard} B.,  2011, A\&A 525, A51

\bibitem[\protect\astroncite{Blandford \& Rees}{1978}]{Blandford1978}
Blandford R.D., Rees M.J.,  1978,
\newblock In: {A.~M.~Wolfe} (ed.) BL Lac Objects., p.328

\bibitem[\protect\astroncite{Bloom}{2008}]{Bloom2008}
Bloom S.D.,  2008, AJ 136, 1533

\bibitem[\protect\astroncite{Cash}{1979}]{Cash1979}
Cash W.,  1979, ApJ 228, 939

\bibitem[\protect\astroncite{Cohen et~al.}{2007}]{Cohen2007}
Cohen M.H., Lister M.L., Homan D.C., et~al., 2007, ApJ 658, 232

\bibitem[\protect\astroncite{Danziger et~al.}{1979}]{Danziger1979}
Danziger I.J., Fosbury R.A.E., Goss W.M., Ekers R.D.,  1979, MNRAS 188, 415

\bibitem[\protect\astroncite{Falomo et~al.}{1994}]{Falomo1994}
Falomo R., Scarpa R., Bersanelli M.,  1994, ApJS 93, 125

\bibitem[\protect\astroncite{{Fossati} et~al.}{1998}]{fossati:98a}
{Fossati} G., {Maraschi} L., {Celotti} A., et~al., 1998, MNRAS 299, 433

\bibitem[\protect\astroncite{Ghisellini et~al.}{2009}]{Ghisellini2009}
Ghisellini G., Maraschi L., Tavecchio F.,  2009, MNRAS 396, L105

\bibitem[\protect\astroncite{{Ghisellini} et~al.}{2010}]{ghisellini:10a}
{Ghisellini} G., {Tavecchio} F., {Foschini} L., et~al., 2010, MNRAS 402, 497

\bibitem[\protect\astroncite{Hartman et~al.}{1992}]{Hartman1992}
Hartman R.C., Bertsch D.L., Fichtel C.E., et~al., 1992, ApJ 385, L1

\bibitem[\protect\astroncite{Healey et~al.}{2008}]{Healey2008}
Healey S.E., Romani R.W., Cotter G., et~al., 2008, ApJS 175, 97

\bibitem[\protect\astroncite{Helene}{1983}]{Helene1983}
Helene O.,  1983, NIMPR 212, 319

\bibitem[\protect\astroncite{Helsel}{2005}]{Helsel2005}
Helsel D.R.,  2005,
\newblock {Nondetects and Data Analysis: Statistics for Censored Environmental
  Data},
\newblock Wiley, New York

\bibitem[\protect\astroncite{Hudson et~al.}{2001}]{Hudson2001}
Hudson M.J., Lucey J.R., Smith R.J., et~al., 2001, MNRAS 327, 265

\bibitem[\protect\astroncite{Impey}{1996}]{Impey1996}
Impey C.,  1996, AJ 112, 2667

\bibitem[\protect\astroncite{Impey \& Tapia}{1990}]{Impey1990}
Impey C.D., Tapia S.,  1990, ApJ 354, 124

\bibitem[\protect\astroncite{Jackson et~al.}{2002}]{Jackson2002}
Jackson C.A., Wall J.V., Shaver P.A., et~al., 2002, A\&A 386, 97

\bibitem[\protect\astroncite{Kellermann \& Pauliny-Toth}{1969}]{Kellermann1969}
Kellermann K.I., Pauliny-Toth I.I.K.,  1969, ApJ 155, L71

\bibitem[\protect\astroncite{Kovalev et~al.}{2009}]{Kovalev2009}
Kovalev Y.Y., Aller H.D., Aller M.F., et~al., 2009, ApJL 696, L17

\bibitem[\protect\astroncite{Kovalev et~al.}{2005}]{Kovalev2005}
Kovalev Y.Y., Kellermann K.I., Lister M.L., et~al., 2005, AJ 130, 2473

\bibitem[\protect\astroncite{{Le{\'o}n-Tavares} et~al.}{2011}]{LeonTavares2011}
{Le{\'o}n-Tavares} J., Valtaoja E., Tornikoski M., et~al., 2011, A\&A 532, A146

\bibitem[\protect\astroncite{Linford et~al.}{2012}]{Linford2012b}
Linford J.D., Taylor G.B., Schinzel F.K.,  2012, ApJ 757, 25

\bibitem[\protect\astroncite{Lister}{2007}]{Lister2007}
Lister M.L.,  2007,
\newblock In: {Ritz} S., {Michelson} P., {Meegan} C.A. (eds.) The First GLAST
  Symposium, Vol. 921. AIPC, p.345

\bibitem[\protect\astroncite{Lister et~al.}{2011}]{Lister2011}
Lister M.L., Aller M., Aller H., et~al., 2011, ApJ 742, 27

\bibitem[\protect\astroncite{Lister et~al.}{2013}]{Lister2013}
Lister M.L., Aller M.F., Aller H.D., et~al., 2013, AJ 146, 120

\bibitem[\protect\astroncite{Lister \& Homan}{2005}]{Lister2005}
Lister M.L., Homan D.C.,  2005, AJ 130, 1389

\bibitem[\protect\astroncite{Lister et~al.}{2009}]{Lister2009_gamma_radio}
Lister M.L., Homan D.C., Kadler M., et~al., 2009, ApJL 696, L22

\bibitem[\protect\astroncite{Lott et~al.}{2012}]{Lott2012}
Lott B., Cavazzuti E., Cutini S., et~al., 2012,
\newblock In: Proc. ``Fermi \& Jansky". eConf C1111101

\bibitem[\protect\astroncite{Lovell et~al.}{2013}]{Lovell2013}
Lovell J.E.J., McCallum J.N., Reid P.B., et~al., 2013, Journal of Geodesy 87,
  527

\bibitem[\protect\astroncite{Mao}{2011}]{Mao2011}
Mao L.S.,  2011, New Astronomy 16, 503

\bibitem[\protect\astroncite{Maraschi et~al.}{1992}]{Maraschi1992}
Maraschi L., Ghisellini G., Celotti A.,  1992, ApJL 397, L5

\bibitem[\protect\astroncite{Massaro et~al.}{2009}]{Massaro2009}
Massaro E., Giommi P., Leto C., et~al., 2009, A\&A 495, 691

\bibitem[\protect\astroncite{Mattox et~al.}{1996}]{Mattox1996}
Mattox J.R., Bertsch D.L., Chiang J., et~al., 1996, ApJ 461, 396

\bibitem[\protect\astroncite{Max-Moerbeck et~al.}{2013}]{Max-Moerbeck2013}
Max-Moerbeck W., Richards J.L., Pavlidou V., et~al., 2013,
\newblock In: 2012 Fermi Symposium proceedings. eConf C121028

\bibitem[\protect\astroncite{Moellenbrock et~al.}{1996}]{Moellenbrock1996}
Moellenbrock G.A., Fujisawa K., Preston R.A., et~al., 1996, AJ 111, 2174

\bibitem[\protect\astroncite{{M{\"u}ller} et~al.}{2014}]{Mueller2014}
{M{\"u}ller} C., {Kadler} M., {Ojha} R., et~al., 2014, A\&A 569, A115

\bibitem[\protect\astroncite{{M{\"u}ller} et~al.}{2011}]{Mueller2011}
{M{\"u}ller} C., Kadler M., Ojha R., et~al., 2011, A\&A 530, L11

\bibitem[\protect\astroncite{Nolan et~al.}{2012}]{2FGL2012}
Nolan P.L., Abdo A.A., Ackermann M., et~al., 2012, ApJS 199, 31 {(2FGL)}

\bibitem[\protect\astroncite{Ojha et~al.}{2004}]{Ojha2004}
Ojha R., Fey A.L., Johnston K.J., et~al., 2004, AJ 127, 1977

\bibitem[\protect\astroncite{Ojha et~al.}{2010}]{Ojha2010}
Ojha R., Kadler M., {B{\"o}ck} M., et~al., 2010, A\&A 519, A45

\bibitem[\protect\astroncite{Padovani \& Giommi}{1995}]{Padovani1995}
Padovani P., Giommi P.,  1995, MNRAS 277, 1477

\bibitem[\protect\astroncite{Pita et~al.}{2014}]{Pita2014}
Pita S., Goldoni P., Boisson C., et~al., 2014, A\&A 565, A12

\bibitem[\protect\astroncite{Pushkarev et~al.}{2010}]{Pushkarev2010}
Pushkarev A.B., Kovalev Y.Y., Lister M.L.,  2010, ApJ 722, L7

\bibitem[\protect\astroncite{Rau et~al.}{2012}]{Rau2012}
Rau A., Schady P., Greiner J., et~al., 2012, A\&A 538, A26

\bibitem[\protect\astroncite{Sbarufatti et~al.}{2009}]{Sbarufatti2009}
Sbarufatti B., Ciprini S., Kotilainen J., et~al., 2009, AJ 137, 337

\bibitem[\protect\astroncite{{Shaw} et~al.}{2013}]{Shaw2013}
{Shaw} M.S., {Filippenko} A.V., {Romani} R.W., et~al., 2013, AJ 146, 127

\bibitem[\protect\astroncite{{Shaw} et~al.}{2012}]{Shaw2012}
{Shaw} M.S., {Romani} R.W., {Cotter} G., et~al., 2012, ApJ 748, 49

\bibitem[\protect\astroncite{{Tavecchio} et~al.}{2010}]{tavecchio:10a}
{Tavecchio} F., {Ghisellini} G., {Bonnoli} G., {Ghirlanda} G.,  2010, MNRAS
  405, L94

\bibitem[\protect\astroncite{Taylor et~al.}{2007}]{Taylor2007}
Taylor G.B., Healey S.E., Helmboldt J.F., et~al., 2007, ApJ 671, 1355

\bibitem[\protect\astroncite{Thompson et~al.}{1993}]{Thompson1993_egret}
Thompson D.J., Bertsch D.L., Fichtel C.E., et~al., 1993, ApJS 86, 629

\bibitem[\protect\astroncite{Tingay \& Edwards}{2002}]{Tingay2002}
Tingay S.J., Edwards P.G.,  2002, AJ 124, 652

\bibitem[\protect\astroncite{Tingay et~al.}{2003}]{Tingay2003a}
Tingay S.J., Jauncey D.L., King E.A., et~al., 2003, PASJ 55, 351

\bibitem[\protect\astroncite{Tingay et~al.}{1998}]{Tingay1998}
Tingay S.J., Murphy D.W., Lovell J.E.J., et~al., 1998, ApJ 497, 594

\bibitem[\protect\astroncite{Tingay et~al.}{2001a}]{Tingay2001_CenA}
Tingay S.J., Preston R.A., Jauncey D.L.,  2001a, AJ 122, 1697

\bibitem[\protect\astroncite{Tingay et~al.}{2001b}]{Tingay2001}
Tingay S.J., Preston R.A., Lister M.L., et~al., 2001b, ApJ 549, L55

\bibitem[\protect\astroncite{Torniainen et~al.}{2008}]{Tornianinen2008}
Torniainen I., Tornikoski M., Turunen M., et~al., 2008, A\&A 482, 483

\bibitem[\protect\astroncite{Tornikoski et~al.}{2001}]{Tornikoski2001}
Tornikoski M., Jussila I., Johansson P., et~al., 2001, AJ 121, 1306

\bibitem[\protect\astroncite{Tzioumis et~al.}{2010}]{Tzioumis2010}
Tzioumis A.K., Tingay S.J., Stansby B., et~al., 2010, AJ 140, 1506

\bibitem[\protect\astroncite{Urry \& Padovani}{1995}]{Urry1995}
Urry C.M., Padovani P.,  1995, PASP 107, 803

\bibitem[\protect\astroncite{{V{\'e}ron-Cetty} \&
  {V{\'e}ron}}{2006}]{Veron2006}
{V{\'e}ron-Cetty} M.P., {V{\'e}ron} P.,  2006, A\&A 455, 773

\bibitem[\protect\astroncite{Wilkes}{1986}]{Wilkes1986}
Wilkes B.J.,  1986, MNRAS 218, 331

\bibitem[\protect\astroncite{Wills et~al.}{2004}]{Wills2004}
Wills K.A., Morganti R., Tadhunter C.N., et~al., 2004, MNRAS 347, 771

\end{thebibliography}
\end{document}